\documentclass[12pt,preprint]{aastex}

\shorttitle{Cuspy triaxial galaxies with dark matter}

\shortauthors{Capuzzo-Dolcetta et al.}

\begin{document}
\def\gap{\;\rlap{\lower 2.5pt
 \hbox{$\sim$}}\raise 1.5pt\hbox{$>$}\;}
\def\lap{\;\rlap{\lower 2.5pt
   \hbox{$\sim$}}\raise 1.5pt\hbox{$<$}\;}

\title{Self-consistent models of cuspy triaxial galaxies with dark matter haloes}

\author{R. Capuzzo-Dolcetta}
\email{roberto.capuzzodolcetta@uniroma1.it}
\affil{Dep. of Physics, University of Roma 'La Sapienza', P.le A. Moro 5,
I-00185, Roma, Italy}
\author{L. Leccese}
\email{linda.leccese@uniroma1.it}
\affil{Dep. of Physics, University of Roma 'La Sapienza', P.le A. Moro 5,
I-00185, Roma, Italy}
\author{D. Merritt}
\email{merritt@astro.rit.edu}
\affil{Dep. of Physics, Rochester Inst. of Technology, 84 Lomb Memorial 
drive, Rochester, NY 14623 - USA}
\author{A. Vicari}
\email{alessandrovicari@uniroma1.it}
\affil{Dep. of Physics, University of Roma 'La Sapienza', P.le A. Moro 5,
I-00185, Roma, Italy}

\begin{abstract}
We have constructed realistic, self-consistent models of
triaxial elliptical galaxies embedded in triaxial dark matter haloes.
We examined three different models for the shape of the dark matter halo:
(i) the same axis ratios as the luminous matter (0.7:0.86:1);
(ii) a more prolate shape (0.5:0.66:1);
(iii) a more oblate shape (0.7:0.93:1).
The models were obtained by means of the standard orbital
superposition technique introduced by Schwarzschild.
Self-consistent solutions were found in each of the three cases.
Chaotic orbits were found to be important in all of the models,
and their presence was shown to imply a possible slow evolution
of  the shapes of the haloes.
Our results demonstrate for the first time that triaxial dark matter
haloes can co-exist with triaxial galaxies.
\end{abstract}


\section{Introduction}
\label{intro}

According to the hierarchical picture of structure formation,
galaxies originated from baryons that condensed within dark-matter haloes.
If this picture is correct, dark matter should exist within and around
galaxies.
In the case of spiral galaxies, this has been widely confirmed by the
study of rotation curves that remain approximately flat beyond
the visible disk, suggesting the presence of invisible matter at
large distances from the galactic centre (Rubin 1991).
In the case of elliptical galaxies, the general absence of
rotating gas disks makes this confirmation more difficult.
Giant elliptical galaxies nevertheless show evidence of
dark matter via stellar kinematics (Kronawitter et al. 2000),
X-ray emission (Mathews \& Brighenti 2003),
and gravitational lensing (Keeton 2001).
Numerical simulations of disk galaxy mergers including dark matter,
like the recent work of Dekel et al. (2005),
produce elliptical galaxies with stellar kinematics
similar to what is observed.

Large-scale simulations of structure formation typically
exclude gas and stars, due to the computational complexities
associated with gaseous dissipation, radiation, star formation, etc.
These simulations make well-defined predictions about the
structure of baryon-free dark matter haloes and the dependence of halo
properties on mass and formation redshift
(Navarro et al. 2004; Diemand, Moore \& Stadel 2004).

Much recent work has focussed on the shapes of the haloes
(Allgood et al. 2006, Macci\'o et al. 2006; Bett et al. 2006).
Simulated haloes are found to be generically prolate/triaxial.
The mean short-to-long axis ratio is $0.5\lap \langle c/a \rangle \lap 0.7$
and the distribution of $c/a$ is approximately Gaussian,
with a sharp cutoff below $c/a\approx 0.3$ that probably
reflects the onset of a dynamical bending instability
(Merritt \& Sellwood 1994).
There are indications that mean halo elongation increases
with halo mass (Allgood et al. 2006).
The shapes of these simulated haloes are apparently supported
mainly by anisotropic velocities rather than by figure
rotation (Warren et al. 1992).

To the extent that luminous elliptical galaxies formed
in a manner analogous to the (simulated) dark-matter haloes --
i.e. through dissipationless clustering of smaller galaxies --
one expects the structure of the two sorts of system to
be similar.
Confirmation of this was recently demonstrated
in the case of radial density profiles
(Merritt et al. 2005a,b):
the same fitting function that best describes luminous
elliptical galaxies is an equally good description of
the $N$-body haloes.
Furthermore the simulated haloes appear
to lie on the same relation between surface brightness
and size (the ``Kormendy relation'') defined by the
galaxies (Graham et al. 2006).

With regard to shapes, observations of elliptical galaxies
yield only the projected isophotes and so determination
of the intrinsic shape distribution requires either
statistical analyses based on large samples
(e.g. Ryden 1992, 1996; Tremblay \& Merritt 1995, 1996;
Ryden, Lauer \& Postman 1993; Alam \& Ryden 2002;
Vincent \& Ryden 2005; Plionis, Basilakos \& Ragone-Figueroa 2006), or
mapping of potentials via detailed kinematical data for individual galaxies
(e.g. Franx, Illingworth \& de Zeeuw 1991; Statler et al. 2004).
Isophotal studies reveal that luminous and faint
elliptical galaxies have significantly different
shape distributions: the latter have $\langle c/a\rangle\approx 0.75$
while the former are flatter, $\langle c/a\rangle\approx 0.65$
(Tremblay \& Merritt 1996).
The apparent shape distributions are found to be mildly
inconsistent with axial symmetry in the sense that too
few, nearly-round galaxies are observed,  but
these studies do not strongly constrain the
degree or distribution of triaxialities.
However, detailed kinematical studies of individual galaxies
sometimes find evidence for significant triaxiality
(Statler et al. 2004).

Still less is known about the true shapes of dark
matter haloes, although constraints come from X-ray
studies, lensing, and Milky Way kinematics
(Tasitsiomi 2003).
When comparing such data to the predictions of the
$N$-body models, it is important to keep in mind
that halo shapes are likely to be influenced by the
presence of the baryons.
For instance, if baryonic matter accumulates at the
center of the dark-matter potential,
the halo is expected to become rounder and less triaxial
(Merritt \& Quinlan 1998; Springel, White \& Hernquist 2004;
Kazantzidis et al. 2004).

Given these uncertainties, it is worthwhile to investigate
new lines of evidence that might constrain the shapes
of dark-matter haloes.
Self-consistency studies constitute one such approach,
which until now has not been explored.
Can stationary, dark-matter haloes of arbitrary shape co-exist with
a galaxy, or does the existence of a central, baryonic
mass concentration imply restrictions on the elongation or
triaxiality of the halo?

In gravitational potentials that respect special symmetries
(spherical- or axi-symmetry, for instance),
isolating integrals of the motion exist in addition to the
energy.
Jeans's (1915) theorem provides a simple prescription for the
construction of steady-state phase-space populations
in such potentials:
the phase-space density must be a function only of the
isolating integrals.
Orbits in triaxial potentials sometimes respect integrals
in addition to the energy (Schwarzschild 1979),
but the mathematical form of these integrals is typically unknown.
Further, many orbits in triaxial potentials
are observed to respect no integrals aside from the energy
(Merritt 1980; Valluri \& Merritt 1998;
Papaphilippou \& Laskar 1998).

Jeans's theorem can be generalized to the non-integrable
case: a steady-state galaxy is one in which the
phase-space density is constant in every connected region
(Kandrup 1998).
An example of a connected region is the invariant torus
of a regular orbit, or the Arnold web corresponding to the
chaotic parts of phase space at a given energy
(Tabor 1989; Rasband 1990).
Howevever Jeans's theorem says nothing about how this
special state is to be achieved, and stochastic orbits
are often observed to remain confined to subsets of the
full allowed region for long periods (e.g. Habib, Kandrup \& Mahon 1997),
making it unlikely that they would achieve a uniform
phase-space population even in a Hubble time.

In the current work, we use the method introduced in 1979 by
Schwarzschild for the numerical construction of stationary
galaxy models.
Schwarzschild's method was tested on a modified Hubble profile (Schwarzschild
1979) and on a Plummer model (Siopis 1999). Later, it was used by
Statler (1987) to demonstrate the self-consistency of model galaxies
made of stars distributed according a triaxial density profile with
a central core.
Subsequently, high resolution observations of
HST (Crane et al. 1993; Ferrarese et al. 1994; Lauer et al. 1995)
showed that almost all the ellipticals have densities that rise
toward the centre approximately as power laws $\rho\sim
r^{-\gamma}$, with $\gamma =2$ for fainter galaxies and $\gamma =1$
for brighter ones. Merritt \& Fridman (1996; hereafter MF96) used
Schwarzschild's method to construct self-consistent solutions
galaxy models with both weak and strong cusps.
In the case of strong cusps, the large number of
stochastic orbits made it more difficult to obtain equilibrium models.
Subsequent studies by Merritt (1997), Siopis (1999) and Terzic (2002)
confirmed and extended the MF96 results.
In these studies, attempts were made to represent the
stochastic orbits via fully-mixed, steady-state distributions
at each energy.

All the works just cited dealt with galaxies containing a
single (luminous) component only.
Here, we introduce a second component,
the dark matter, presumably consisting of long-lived,
``cold,'' and collisionless particles,
e.g. supersymmetric neutralinos (Bertone, Hooper \& Silk 2004).

In the following Section we explain the
method -- a generalization of Schwarzschild's -- for constructing
self-consistent equilibrium models.
The details of these models are discussed in Section $\ref{modello}$ while
in Section $\ref{catalogo}$ we describe their orbital characteristics. In
Section $\ref{risultati}$ our results are analyzed and compared to
those of models without dark matter.
Finally in S ection $\ref{discussione}$ we discuss the consequences of
stochasticity for the long-term stability of the models.

\section{Method}\label{autoconsistenza}
In the absence of analytic expressions for the orbital integrals,
one must resort to numerical methods to construct self-consistent
models of galaxies or dark-matter haloes.
To obtain a numerical solution to the problem we follow the
Schwarzschild (1979) scheme. The procedure can be summarized as
follows:

(1) A smooth, three-dimensional mass distribution resembling
observed galaxies is postulated and the corresponding gravitational potential
calculated by means of Poisson's equation.

(2) A library of orbits is realized by integration of the
equations of motion in this potential starting from a large
set of different initial conditions.

(3) The configuration space is divided into cells and
the time each orbit spends in any cell is recorded.

(4) A linear combination of these orbits (with non-negative weights)
is sought which gives the best discrete approximation to the
known masses of the cells.

We generalized Schwarzschild's method to the construction of
two-component (luminous and dark matter), self-consistent models
as follows.
First, we constructed a catalog of orbits integrating the equations
of motion for a particle moving in the potential generated by the sum
of the luminous and dark matter density distributions.
Then, two different grids of cells, one for the luminous and one for the
dark matter components, were built.
The technique used to realize each grid is the same described in
MF96:
the space was first divided by $21$ concentric ellipsoidal
shells in zones containing an equal amount of one of the two matter
components.
Then, each octant was divided into three sectors corresponding to
 \begin{itemize}
 \item[(i)] the volume with $ax>by$ and $ax>cz$,
 \item[(ii)] the volume with $by>ax$ and $by>cz$,
 \item[(iii)]the volume with $cz>ax$ and $cz>by$.
\end{itemize}
Finally, we used a set of six planes to divide each sector in $16$
regions.
In this way, we obtained a total of $1008$ cells per octant for each grid.
The fraction of time spent by each orbit of this catalog in both
sets of cells was recorded.
We then used the data for the luminous-matter grid
(called \emph{Luminous}) to reproduce, by
means of a linear superposition of orbits with non-negative
occupation numbers, the given mass of the luminous component in each
cell of that grid.
The same procedure was followed to reproduce the
dark matter distribution on the \emph{Dark matter} grid.

In practice, our optimization problem consisted of minimizing,
separately, the quantities
\begin{equation}\label{autclum}
\chi_{lum}^{2}=\frac{1}{N_{lum}} \sum_{j=1}^{N_{lum}}
\Big(M_{j;lum}-\sum_{k=1}^{n_{orb}} C_{k;lum} B_{k,j;lum} \Big)^2
\end{equation}
and
\begin{equation}\label{autcdm}
\chi_{dm}^{2}=\frac{1}{N_{dm}} \sum_{j=1}^{N_{dm}}
\Big(M_{j;dm}-\sum_{k=1}^{n_{orb}} C_{k;dm} B_{k,j;dm} \Big)^2
\end{equation}
where $B_{k,j;lum}$ is the fraction of time that the \emph{k}th
orbit spends in the \emph{j}th cell of the \emph{Luminous} grid and
$B_{k,j;dm}$ is the same quantity in the \emph{j}th cell
of the \emph{Dark matter} grid; $M_{j;lum}$ is the mass which the
model places in the \emph{j}th cell of the \emph{Luminous} grid and
$M_{j;dm}$ that placed in the \emph{j}th cell of the \emph{Dark
matter} one; $C_{k;lum}$ and $C_{k;dm}$ are the quantities to be
determined by the minimization procedure. The latter represent the
total mass of stars and dark matter particles, respectively, spread
along the \emph{k}th orbit ($1\leq \emph{k} \leq n_{orb}$).

To guarantee non-negative orbital weights,
we imposed the constraints $C_{k;lum}\geq 0$ and $C_{k;dm}\geq 0$.
The NAG routine E04NCF (Stoer 1971; Gill et al. 1984) was used to
solve these constrained least-squares problems.

\section[]{Mass models}\label{modello}
\subsection{Density and potential}
We considered three different mass models for the galaxy+halo
system.
In each of the three cases, we adopted triaxial shapes for both
components but with a different radial distribution for the dark
and luminous matter.
In the first model the dark matter was assumed to have the same axial
ratios as the luminous matter, while
the second and third models were characterized by
haloes that were more prolate and more oblate, respectively,
than the luminous matter.

\begin{figure}
\begin{center}
\includegraphics[width=5in,height=8.5in]{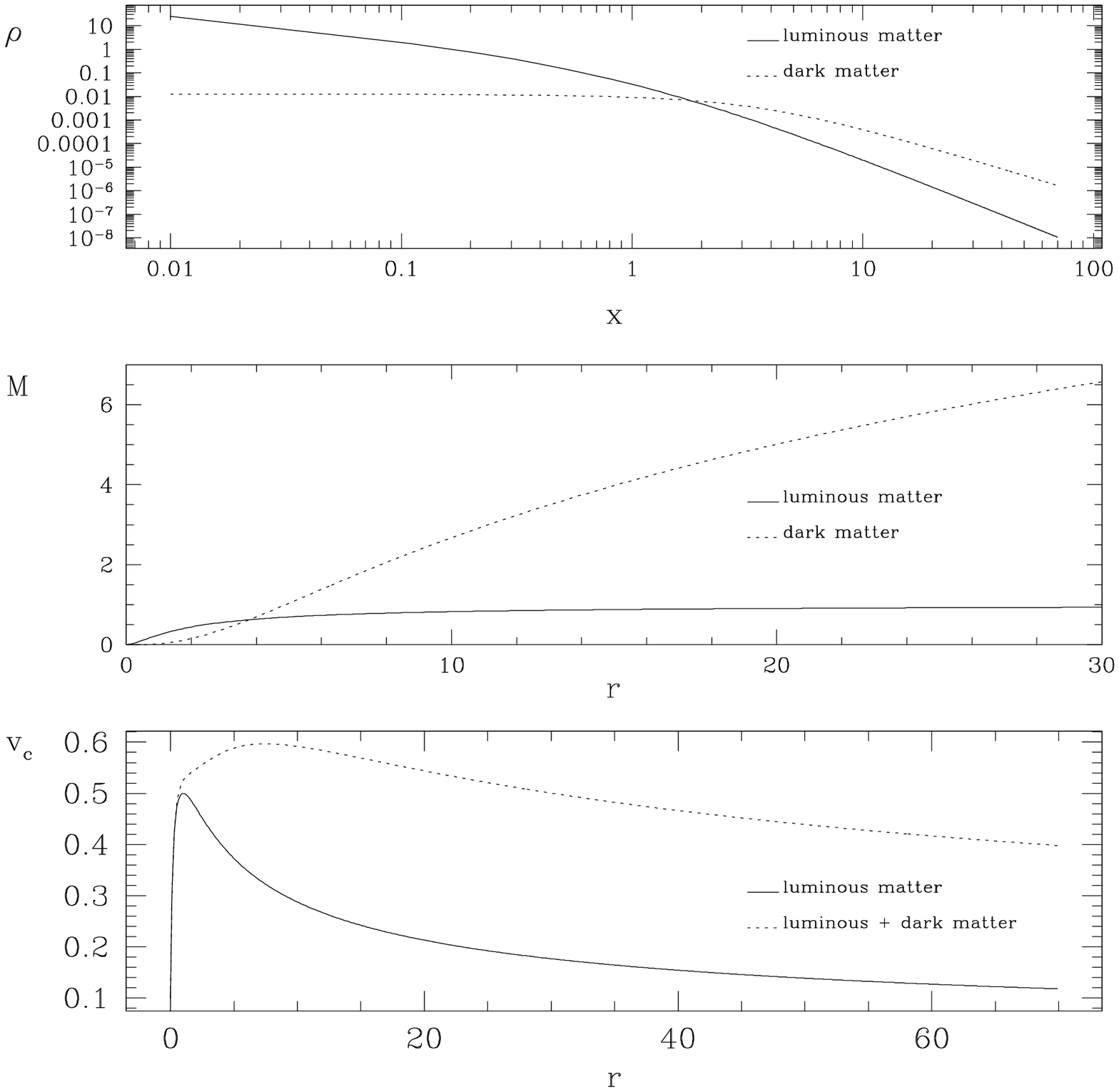}
\caption{Top: long-axis density profiles of the luminous and
dark matter components.
Middle: mass enclosed within spheres of radius $r$ in
the case $a=b=c$.
Bottom: circular velocity profiles in the spherical model,
including both components (dashed line),
and in a model without dark matter. }\label{mixdmv}
\end{center}
\end{figure}

As in MF96, we adopted the following mass distribution for the
luminous component:
\begin{equation}
\label{rhol} \displaystyle \rho_{l}(m) =\frac{M}{2 \pi
a_{l}b_{l}c_{l}} \frac{1}{m (1+m)^3}
\end{equation}
with
\begin{equation}
\label{m}\displaystyle
m^2=\frac{x^{2}}{a_{l}^{2}}+\frac{y^{2}}{b_{l}^{2}}+\frac{z^{2}}{c_{l}^{2}}
\qquad 0 < c_{l}< b_{l}< a_{l}
\end{equation}
and $M$ the total luminous mass (see Fig. \ref{mixdmv}) .

For the dark matter component we adopted
\begin{equation}
\label{rhodm}\displaystyle
\mathop{\rho_{dm}}(m') =\frac{\rho_{dm,0}}{\left(1+m'\right)\left(1+m'^2\right)}
\end{equation}
where
\begin{equation}
\label{m1}\displaystyle
m'^2=\frac{x^{2}}{a_{dm}^{2}}+\frac{y^{2}}{b_{dm}^{2}}+\frac{z^{2}}{c_{dm}^{2}}
\qquad 0 < c_{dm}< b_{dm}< a_{dm},
\end{equation}
and $\rho_{dm,0}$ is the central dark matter density (see Fig.
\ref{mixdmv}).

The density profile in equation~(\ref{rhodm}) was first proposed
by Burkert (1995) for the dark matter haloes of dwarf galaxies, and
later extended to the whole family of spiral (Salucci \& Burkert
2000) and to massive elliptical (Borriello, Salucci \&
Danese 2003, hereafter BSD) galaxies.
By adopting the Burkert profile -- which has a large, low-density core --
we are favoring data over theory,
since $N$-body simulations of gravitational clustering
imply dark-matter densities that increase roughly as a power law  inside
of the halo virial radius (Navarro, Frenk \& White 1996;
Moore et al. 1998; Merritt et al. 2005b).
Rotation curve studies, on the other hand,
are generally interpreted as implying low
($\sim (1-5)\times 10^{-2}M_\odot {\rm pc}^{-3}$)
dark matter densities
within the region ($r< 10^2$ pc) where the rotation
curves are sampled (e.g. Burkert 1995; Salucci \& Burkert 2000;
de Blok \& Bosma 2002; Gentile et al. 2005).
It was to model these low observed dark matter densities that the
Burkert model (\ref{rhodm}) was postulated.

At large radii, the Burkert profile is similar to the
NFW and Moore profiles that are commonly fit to $N$-body
haloes, i.e. $\rho_{dm}\propto r^{-3}$.
In the central regions where the two sorts of profile
(core vs. power law) differ, the gravitational potential
is dominated by the luminous galaxy.
Thus, our  conclusions about  the constraints that a central
galaxy places on the shape of the halo are probably
applicable also to halo models without cores.

The gravitational potential at a point $\textbf{x}=(x,y,z)$ due to a
mass distribution $\rho = \rho (w)$,
where $w^2=x^2/a^2 + y^2/b^2 + z^2/c^2$ may
be written (Chandrasekhar 1969)

\begin{equation}
\label{pottriax}\displaystyle
\Phi(\textbf{x})=-\pi Gabc\int_{0}^{\infty}\frac{[\Psi(\infty)-\Psi(w)]d\tau}{\sqrt{(\tau +a^{2})(\tau +b^{2})(\tau +c^{2})}}
\end{equation}
with
\begin{equation}
\label{psim}\displaystyle
\Psi(w)=2\int_{0}^{w} w'\rho (w')dw'
\end{equation}
and
\begin{equation}
\label{mtau}\displaystyle
w^{2}(\tau)=\frac{x^{2}}{a^{2}+\tau}+\frac{y^{2}}{b^{2}+\tau}+\frac{z^{2}}{c^{2}+\tau}.
\end{equation}
Consequently, we find, for the luminous and dark matter potentials,
respectively,

\begin{equation}
\label{potl} \Phi_{l}(\textbf{x})= \displaystyle -\frac{GM}{2}
\int_{0}^{\infty}\frac{d\tau}{\Delta} \frac{1}{(1+m)^2},
\end{equation}
\begin{eqnarray}
\label{potdm}\displaystyle \lefteqn{ \Phi_{dm}(\textbf{x})=-\pi
Ga_{dm}b_{dm}c_{dm}\rho_{dm,0} {}} \nonumber\\ && \! \! \!
\!\!\!\times\!\! \int_{0}^{\infty} \!\! \frac{d\tau}{\Delta '} \!
\left[ \!\frac{\pi}{2}\!-\!\arctan m'\!+\!\log\!
\left(\!1\!\!+\!\!m'\!\right)\!-\!\frac{1}{2}\!\log
\!\left(\!1\!\!+\!\! m'^{2}\! \right)\right]\!, {}
\end{eqnarray}
with
\begin{equation}
\label{delta}\displaystyle
\qquad\Delta =\sqrt{(\tau +a_l^{2})(\tau +b_l^{2})(\tau +c_l^{2})}
\end{equation}
and
\begin{equation}
\label{delta1}\displaystyle
\qquad \Delta ' =\sqrt{(\tau +a_{dm}^{2})(\tau +b_{dm}^{2})(\tau +c_{dm}^{2})}.
\end{equation}
The components of the gravitational forces are
\begin{equation}
\label{forcel}\displaystyle F_{l_{i}}=-\frac{\partial
\Phi_{l}}{\partial
x_{i}}=-GMx_{i}\int_{0}^{\infty}\frac{d\tau}{\Delta(a_{i}^{2}+\tau)(1+m)^{3}m}
\end{equation}
\begin{eqnarray}
\label{forcedm}\displaystyle
\lefteqn{ F_{dm_{i}}=-\frac{\partial
\Phi_{dm}}{\partial x_{i}}=-2\pi Ga_{dm}b_{dm}c_{dm}\rho_{dm,0}
x_{i}{} }\nonumber\\ & & \quad \quad \times \!\!\int_{0}^{\infty}\!\!\frac{d\tau}{\Delta'
({a'_{i}}^{2}+\tau)(1+m')(1+m'^{2})}
\end{eqnarray}
with $a_{1}=a_l$, $a_{2}=b_l$, $a_{3}=c_l$ ,$a'_{1}=a_{dm}$,
$a'_{2}=b_{dm}$, $a'_{3}=c_{dm}$ and $i=1,2,3$.
These integrals were rewritten using an appropriate
change of variables (see Appendix \ref{potcons})
in order to reduce the complexity of their
numerical treatment.

The mass contained in the ellipsoid of semi-axes $wa$, $wb$ and $wc$ is
\begin{equation}\label{mw}
\displaystyle M(w)=4\pi abc \int_{0}^{w} \rho(w')w'^{2}dw';
\end{equation}
this implies that the total mass is $M_{tot}=\lim_{w \rightarrow
\infty} M(w)$. In the case of luminous and dark matter,
respectively, equation~(\ref{mw}) gives (see Fig. \ref{mixdmv}):
\begin{equation}\label{massl}
\displaystyle
M_{l}(m)=M\left(\frac{m}{1+m}\right)^2,
\end{equation}
\begin{equation}\label{massdm}
\displaystyle
M_{dm}(m')=\pi a_{dm}b_{dm}c_{dm}\rho_{dm,0}B(m'),
\end{equation}
where $B$ is the function
\begin{equation}
\displaystyle\label{funcb} B(m')= -2 \textrm{atan} \left(m'\right)+2
\log \left(1+m'\right) +\log \left(1+ m'^{2} \right) .
\end{equation}
We note that the total mass of the dark matter model is
infinite.
We cut off the dark matter grid at $m=m'_{max}$,
where $m'_{max}\approx 19$ was determined by the requirement
that the mass in dark matter was ten times greater than
the luminous total mass.

\subsection{Model parameters and constraints}
The free parameters of our models are:
\begin{itemize}
\item[-] $a_l$, $b_l$, $c_l$ and $M$ for the luminous component;
\item[-] $a_{dm}$, $b_{dm}$, $c_{dm}$ and $\rho_{dm,0}$ for the dark
matter.
\end{itemize}
With regard to the axis ratios of the luminous component,
there exist some constraints in the form of the observed
distribution of galaxy isophotal shapes
(Ryden 1992; Tremblay \& Merritt 1995, 1996; Alam \& Ryden 2002; Plionis, Basilakos
\& Ragone-Figueroa 2006).
Intrinsic, short-to-long axis ratios of bright ($M_B < -20$) elliptical galaxies
appear to be narrowly clustered around $3/4$ (Tremblay \& Merritt 1996);
the constraints on the second axis ratio are less strong, but most studies
(e.g. Tremblay \& Merritt 1995) find that axisymmetric shapes can be securely ruled out.
We accordingly fixed the short-to-long axis ratio of the luminous
component to be $0.7$ in all of the models and assumed ``maximal triaxiality,''
i.e.
\begin{equation}\label{rapp}
\displaystyle
\frac{c_l}{a_l}=0.7,
\qquad T\equiv \frac{{a_l}^{2}-{b_l}^{2}}{{a_l}^{2}-{c_l}^{2}}=\frac{1}{2}.
\end{equation}

For the dark matter, we investigated three choices for the axis ratios:
equal to those of the luminous matter (MOD1);
a more prolate shape ($T=0.75$) with $c_{dm}/a_{dm}=0.5$ (MOD2);
and a more oblate shape ($T=0.25$) with $c_{dm}/a_{dm}=0.7$ (MOD3)
The parameters of the models are summarized in Table \ref{tab1}.

 \begin{table} \small
 \begin{center}
\begin{tabular}{lcccc}
\multicolumn{5}{c}{\textbf{TABLE 1}} \\
  \hline
  \hline
 \multicolumn{1}{l}{MODEL} & \multicolumn{2}{c}{Luminous matter}& \multicolumn{2}{c}{Dark matter}
  \\
  \multicolumn{1}{l}{} & \multicolumn{1}{c}{$c_{l}/a_{l}$}& \multicolumn{1}{c}{$T_{l}$} &\multicolumn{1}{c}{$c_{dm}/a_{dm}$}& \multicolumn{1}{c}{$T_{dm}$}\\
  \hline
 \multicolumn{1}{l}{MOD1} & \multicolumn{1}{c}{$0.7$}& \multicolumn{1}{c}{$0.5$} &\multicolumn{1}{c}{$0.7$}& \multicolumn{1}{c}{$0.5$}\\
  \hline
   \multicolumn{1}{l}{MOD2} & \multicolumn{1}{c}{$0.7$}& \multicolumn{1}{c}{$0.5$} &\multicolumn{1}{c}{$0.5$}& \multicolumn{1}{c}{$0.75$}\\
  \hline
   \multicolumn{1}{l}{MOD3} & \multicolumn{1}{c}{$0.7$}& \multicolumn{1}{c}{$0.5$} &\multicolumn{1}{c}{$0.7$}& \multicolumn{1}{c}{$0.25$}\\
  \hline

\end{tabular}
 \caption{Axial ratios and triaxial parameter of the density distributions of the 3 
 galactic models.}\label{tab1}
\end{center}
\end{table}

If the unit of length is taken to be $a_l$, there remain two parameters:
the ratio of luminous to dark-matter scale lengths and the normalization
of the dark matter density.


We applied the results of the fundamental plane study of Borriello, Salucci \&
Danese (2003) to the triaxial case, yielding $a_{dm}/a_l=3.64$ and the value of
$\rho_{dm,0}$ corresponding to
\begin{equation}\label{gammae}
\Gamma_{e} \equiv \frac{\displaystyle M_{dm}(m'_e)} {\displaystyle
M_{l}(m_e)} \simeq 0.3
\end{equation}
which is obtained by the evaluation of the luminous matter within the 
ellipsoid delimited by $m_e$, that characterizes the \lq luminous\rq ~
ellipsoid having the $x$-semiaxis equal to the effective radius $R_e=1.81a_l$ 
($m_e$ is the triaxial generalization of the effective radius $R_e$), 
and that of the dark matter contained in the $m'_e=0.5$ ellipsoid
having the same long-axis extension as the luminous one.
According to BSD this ($a_{dm}/a_l,\Gamma_{e}$) pair of values
gives the best fit of theoretical data to the FP. From this
constraint it follows that $\rho_{dm,0}=1.26 \times
10^{-2}M_{l,T}/{a_l}^3$ for MOD1, $\rho_{dm,0}=2.26 \times
10^{-2}M_{l,T}/{a_l}^3$ for MOD2 and $\rho_{dm,0}=1.14 \times
10^{-2}M_{l,T}/{a_l}^3$ for MOD3.


Hereafter we express all quantities in units such that
$G$, $a_l$ and $M$ are equal to unity.
The implied unit of time is
\begin{equation}\label{unitatempo}
T_u=G^{-1/2}{a_l}^{3/2}M^{-1/2}
= 1.49 \times 10^6 
{\rm yr}\Big(\frac{M}{10^{11}M_{\odot}}\Big)^{-1/2}\Big(\frac{a_l}{1 {\rm kpc}}\Big)^{3/2}{}
\end{equation}
and $V_u=\sqrt{GM/a_l}$ is the unit of velocity; the, derived,
energy unit is $E_u=(GM^2/a_l)$.

\section[]{Orbital catalogs}
\label{catalogo}

\subsection{Initial conditions}
The orbital catalogs were built as in Schwarzschild (1993).
Initial conditions consisted either of zero initial velocity
(``stationary'' start-space), or points in the $X-Z$ plane with
$v_x=v_z=0$ and $v_y=\sqrt{2(E-\Phi)} \neq 0$ ($X-Z$ start-space).
As argued by Schwarzschild (1993), these choices cover most of the orbital
types in the full phase space of a nonrotating triaxial model;
in particular, orbits starting in the $X-Z$ plane are mostly
tubes and avoid a region around the center of the model,
while  orbits starting on an equipotential surface are either stochastic or
regular boxlets, that approach the origin after a sufficiently long time.
For each galactic model we fixed $32$ energy values and for each value we
selected a set of $150$ initial conditions
from the $X-Z$ start-space and a set of $192$ initial conditions
from the stationary start-space, thus obtaining a catalog of $10944$
orbits. Each orbit was integrated in time using a $7/8$ order
Runge-Kutta-Fehlberg algorithm with a variable step size so as to
have a relative error in energy less than $10^{-6}$ per time step.

\subsection{Integration times}
Particular care must be given to the choice of the orbital
integration time. Because our aim is to construct a stationary model
of a galaxy, all orbits must be integrated over a time interval long
enough to guarantee a steady-state galaxy model. This means that the
occupation numbers $B_{k,j}$ of each orbit in the grid cells must be
time-invariant. The time required by the $B_{k,j}$ to converge to
stationary values varies from one orbit to another, whether the
orbit is regular or stochastic. In fact, a regular orbit can
fill its invariant torus in a short time or, in the case of a nearly
resonant orbit, it can take much longer to fill it (Merritt \& Valluri 1999).
In the same way, some chaotic orbits diffuse rapidly in the entire
phase-space permitted by their energy, while others remain confined
between regular tori for a very long time (Valluri \& Merritt 2000)
(we will call \emph{ergodic} the orbits with convergent $B_{k,j}$).

To select the integration time for the orbits we followed an iterative
procedure, proposed by Pfenniger (1984), that stops the integration
when the relative variations of $B_{k,j;lum}$ and $B_{k,j;dm}$,
in a time step go below $0.5 \%$.
In detail, we proceeded as follows:
\begin{itemize}
\item[1)] Integrate and store the $B_{k,j;lum}$ and $B_{k,j;dm}$
over a time $\tau = T_d /2$ where $T_d$ is the ``dynamical time,"
defined for each given energy as the period of the $1:1$ resonant
orbit in the $X-Y$ plane (see MF96);
\item[2)] Restart the integration for another time $\tau$,
and store the new $B_{k,j;lum}$ and $B_{k,j;dm}$ at the end;
\item[3)] Compare the new and old $B_{k,j;lum}$ and $B_{k,j;dm}$.
If the maximum relative difference is greater than $0.005$ then
average the two sets of $B_{k,j;lum}$ and $B_{k,j;dm}$,
double $\tau$ and repeat step 2, otherwise stop the computation.
\end{itemize}
As argued by Pfenniger, this method does not converge for all
orbits (e.g. ``sticky'' chaotic orbits), so it is needed to
fix a maximum time of integration, $t_{max}$.


We set this value at $t_{max}=2 T_H$, where $T_H$ is the Hubble time;
the choice $T_H=15 Gyr$ means $t_{max}=10,000T_u$, where $T_u$ is the time
unit in eq. \ref{unitatempo} once assumed $M=10^{11}$ M$_\odot$ and
$a_l=1kpc$.


Most of the CPU time was associated with the integration of the
orbit libraries. Since each orbit is independent, this task is
efficiently parallelized. We used the Plexus cluster at the
Rochester Institute of Technology to carry out the integrations.
The construction of the complete library of 10944 orbits for MOD1
required approximately 6 hours using in parallel 16 nodes
of the cluster.

\section[]{Results}
\label{risultati}
\subsection{Self-consistency}

\begin{figure}
\begin{center}
$\begin{array}{c}
  \includegraphics[width=3.5in,height=3in,angle=0]{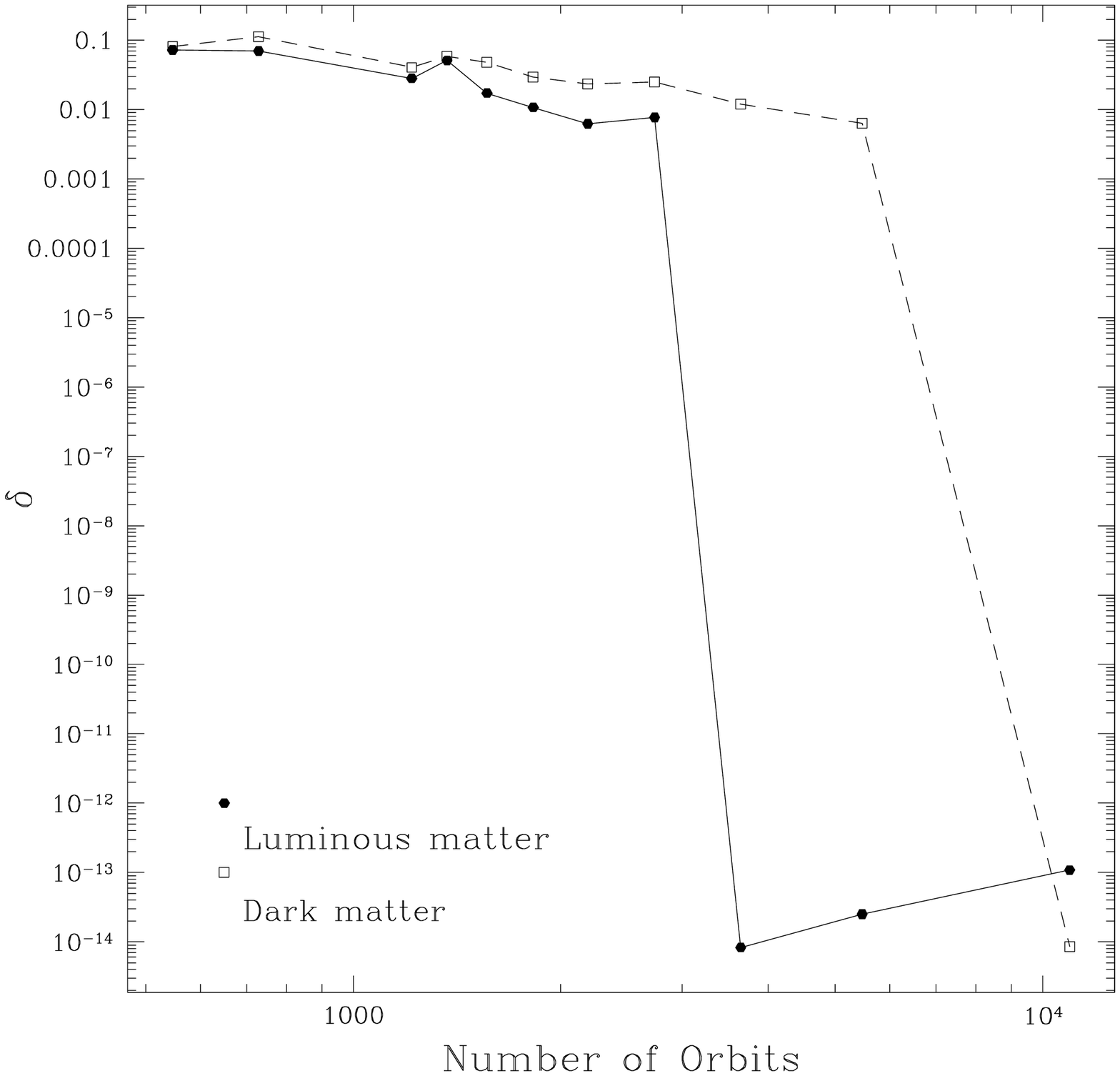} \\
  \includegraphics[width=3.5in,height=3in,angle=0]{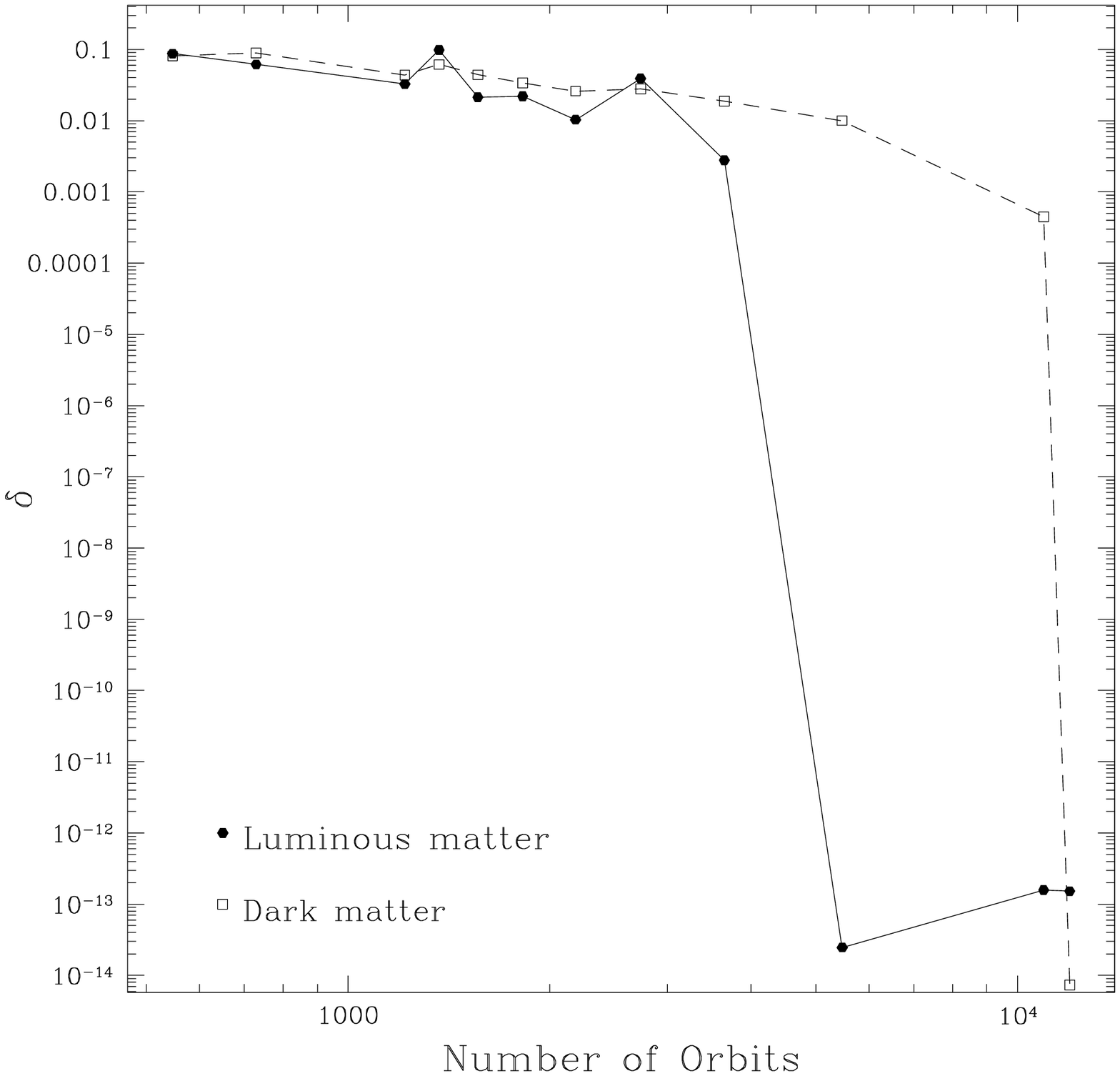} \\
  \includegraphics[width=3.5in,height=3in,angle=0]{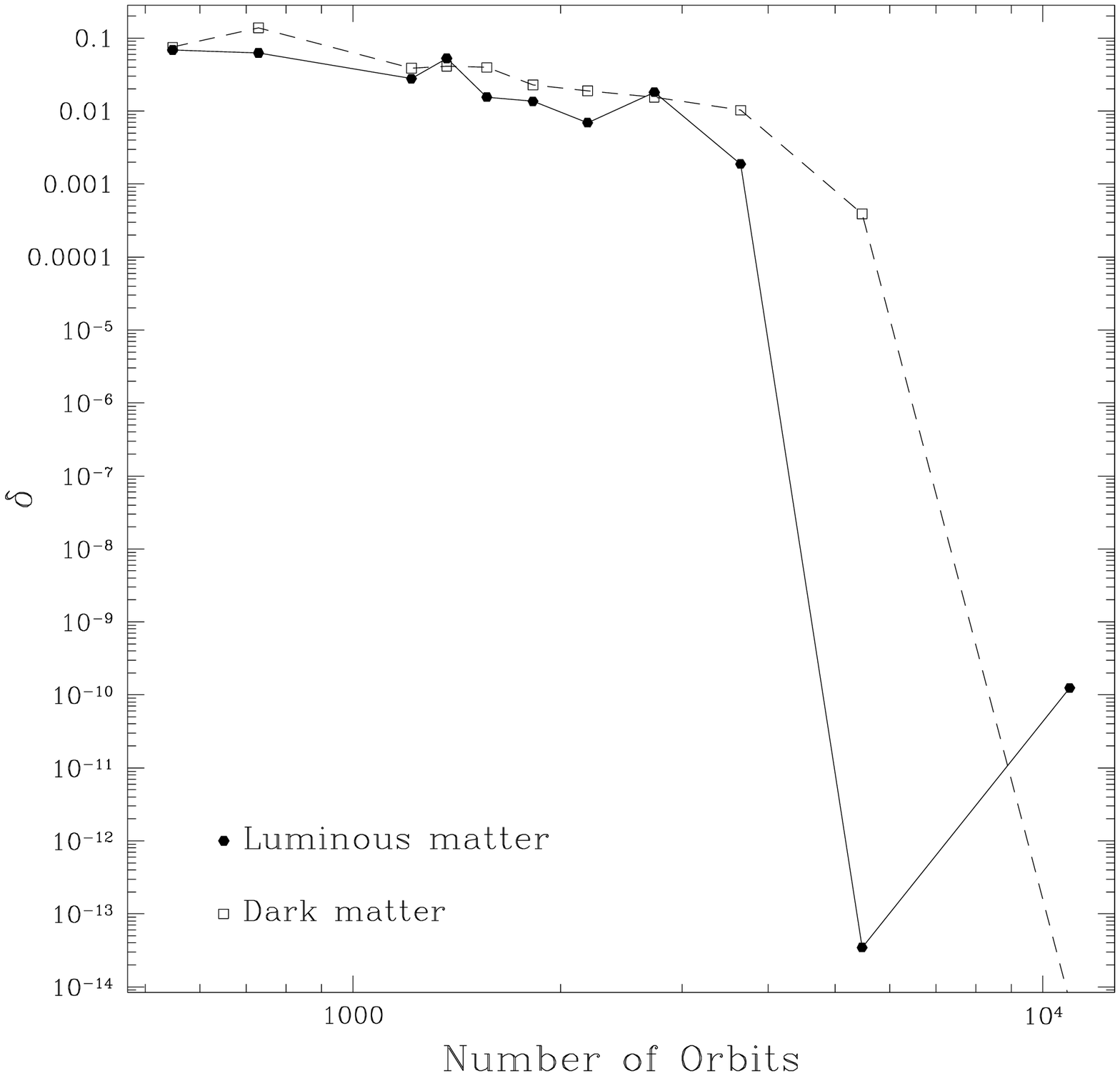}
\end{array}$
\caption{The rms deviation $\delta$
(equation \ref{deltprol}) as a function of the number of orbits
supplied to the optimization routine for (from top to bottom)
MOD1, MOD2 and MOD3,
respectively.}\label{deltagraf}
\end{center}
\end{figure}

\begin{figure}
\begin{center}
\includegraphics[width=6in,height=8.5in]{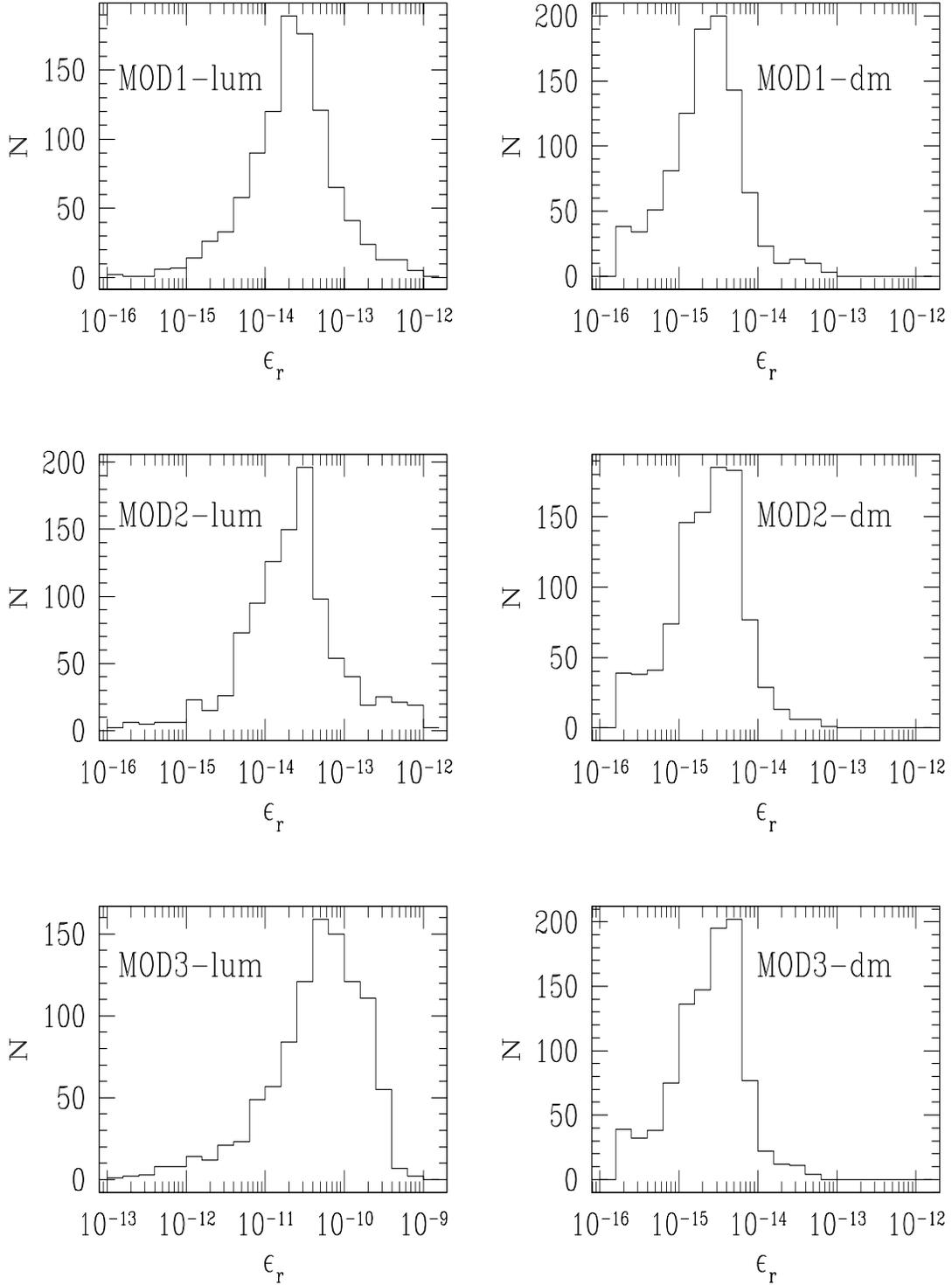}
\caption{Histograms of the relative errors $\epsilon_{r}$ (see text).
They refer to both the components grids of the three models analyzed
and are calculated using the $C_{k}$ quantities of the
self-consistent solutions.} \label{relerr}
\end{center}
\end{figure}

The main result of this work is that we were able to succesfully construct
self-consistent models in all of the cases considered for the shape of
the DM halo.
To obtain the model with the prolate shape of the dark
matter distribution (MOD3) we needed to enlarge the catalog of orbits to
$11970$.
For each matter component, a parameter, $\delta$,
indicating the departure from self-consistency may be defined as
\begin{equation}
 \delta= \frac{\sqrt{\chi ^2}}{\overline{M}}
\label{deltprol}
\end{equation}
where $\overline{M}$ represents the average mass contained in the
grid cells and $\chi^2$ is the quantity defined in Section
\ref{autoconsistenza}. The dependence of $\delta$ on the choice of 
the number of orbits in the catalog is shown in Fig. \ref{deltagraf} for MOD2;
Fig. \ref{relerr}
gives the distribution of the relative error in mass,
$\epsilon$, defined as $\epsilon= |M_{j}-\sum_{k=1}^{n_{orb}} C_{k} B_{k,j}|/
M_{j}$ for each $j$'th grid cell in each of the three self-consistent
models.

If we analyze the orbits that contribute significantly to the self-consistent
solutions, we find that a large fraction consist of orbits that do
not conserve the sign of the components of the angular momentum
(corresponding to regular box orbits and boxlets or to
chaotic orbits). Regarding orbits that conserve the $z-$ component
(named short-axis tubes) or the $x-$ component (outer tubes) of
the angular momentum, it is noted that, to guarantee the prolate
structure of the dark matter in MOD2, the fraction of outer orbits
was greater than that of the short tubes, while the opposite was true
in MOD3, the more oblate model.

\begin{figure}
$\begin{array}{cc}
                \includegraphics[width=3in,height=3in]{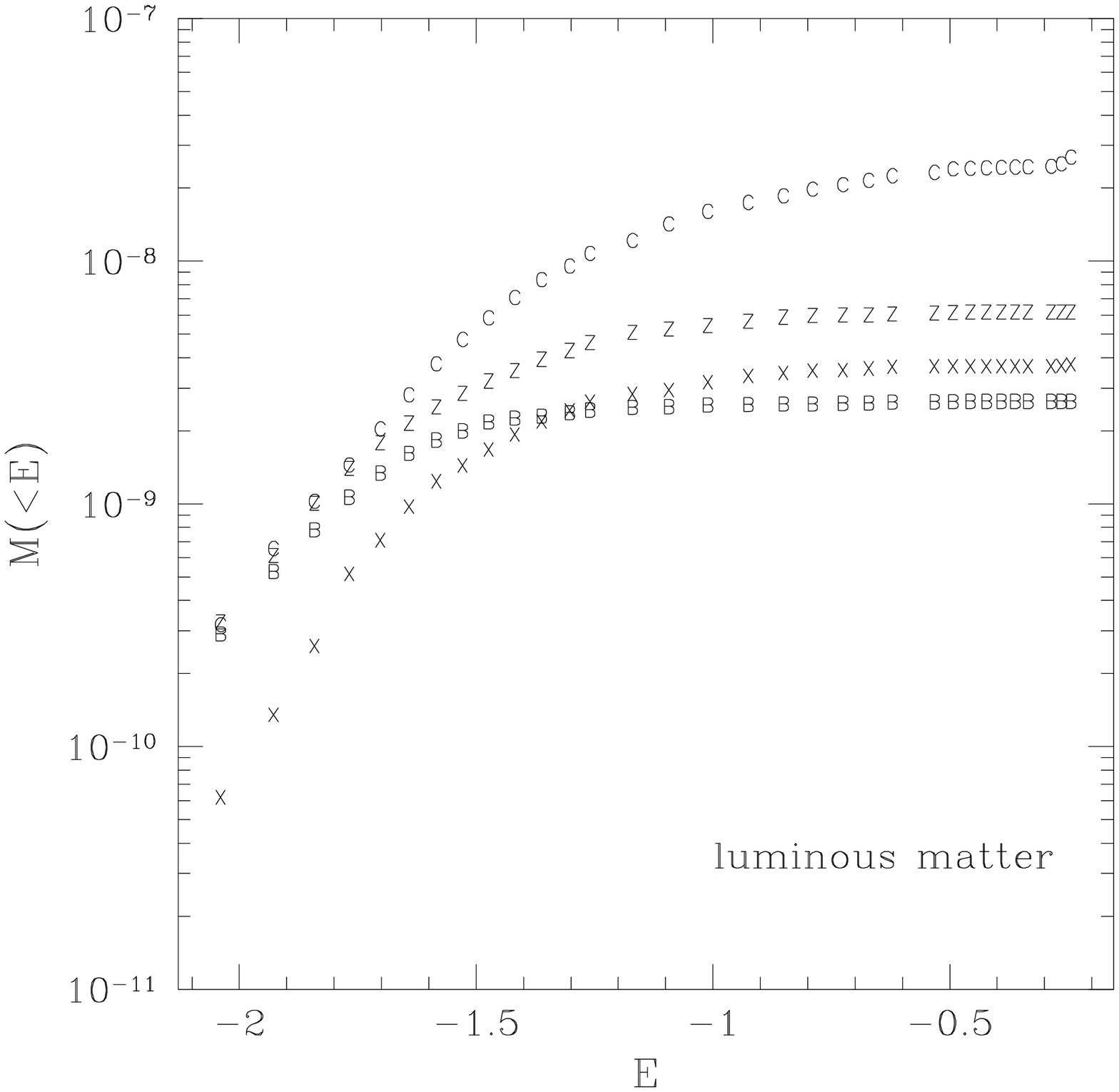}  &
 \includegraphics[width=3in,height=3in]{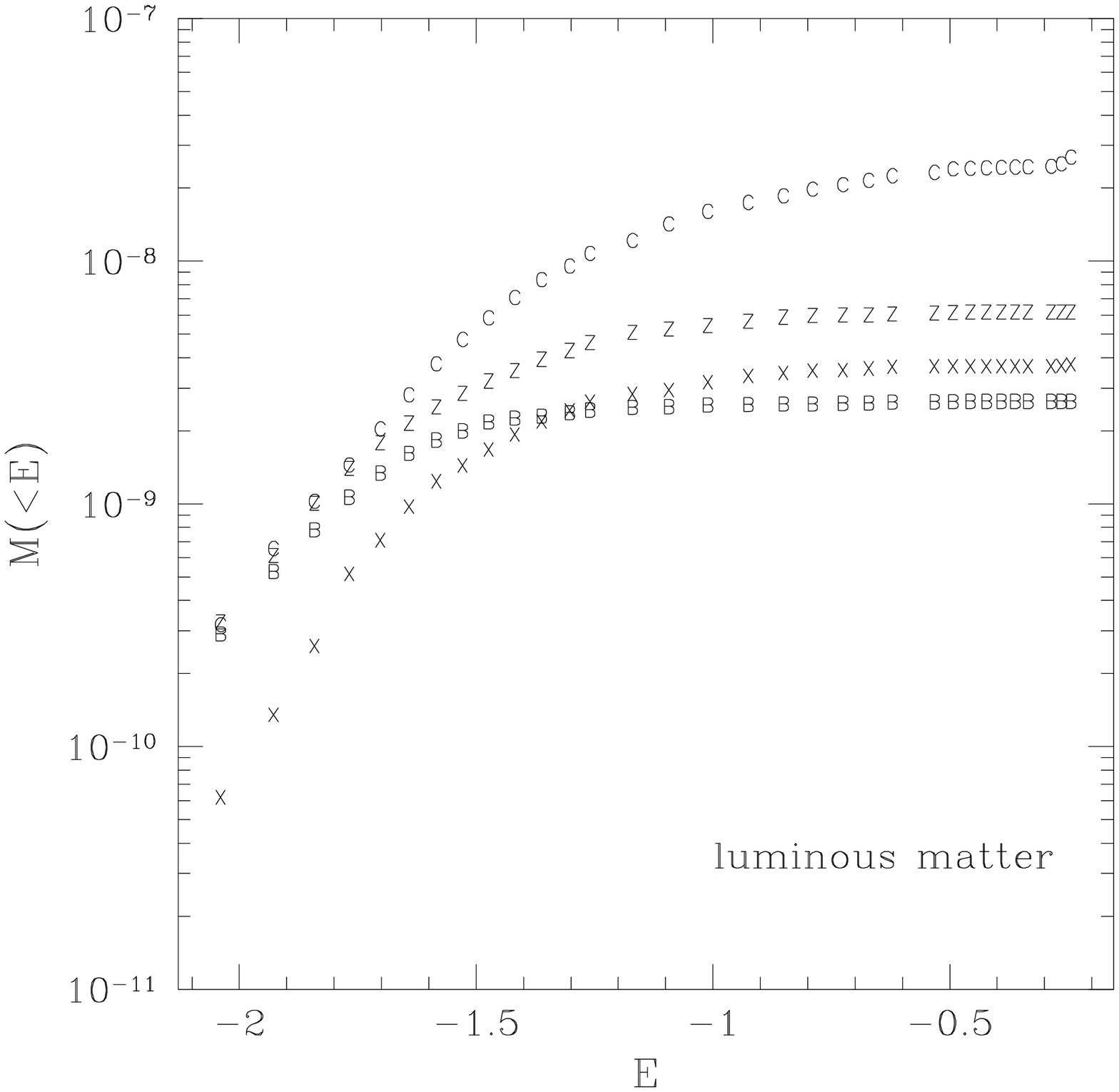} \\
                 \includegraphics[width=3in,height=3in]{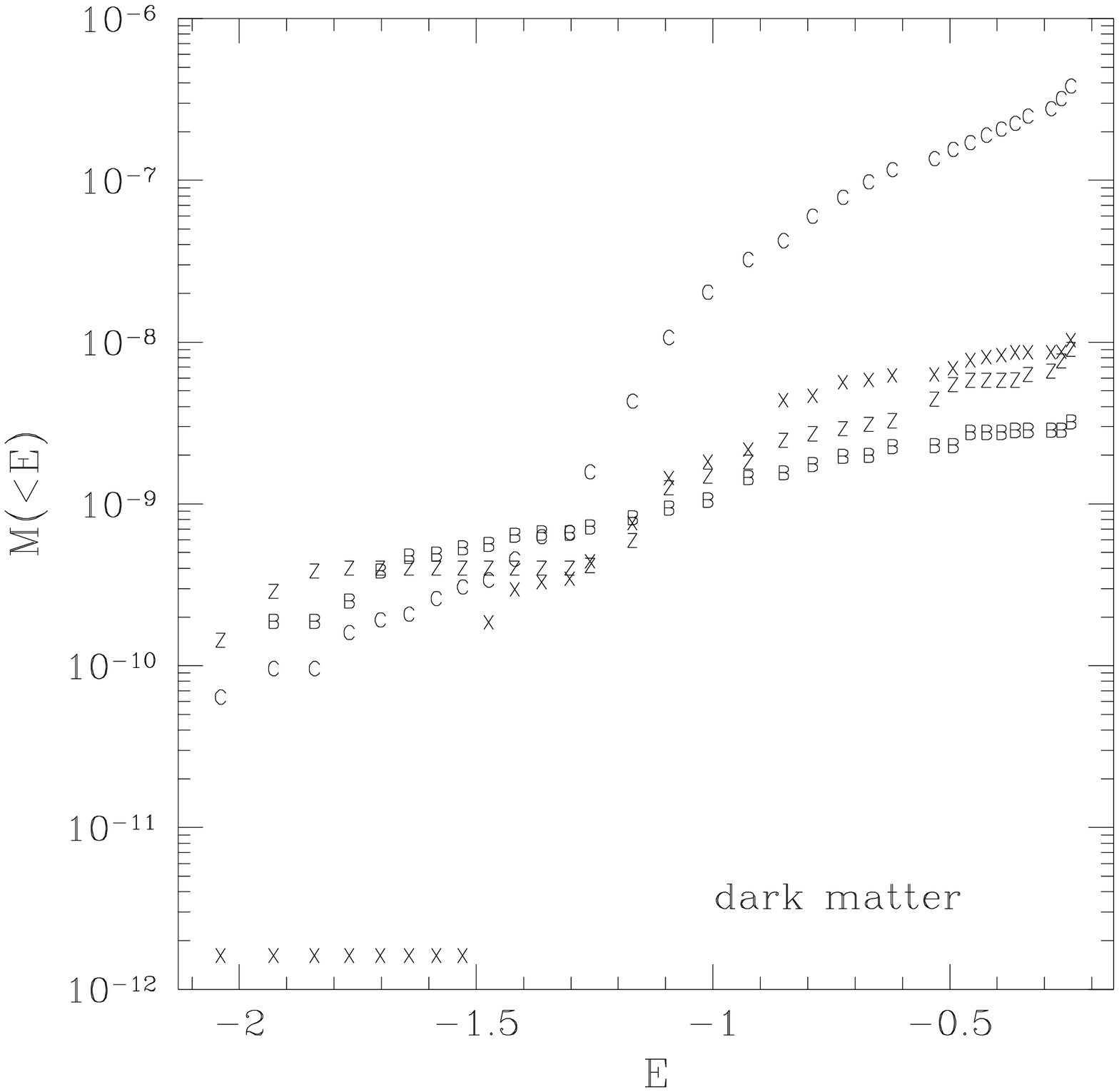} &
\includegraphics[width=3in,height=3in]{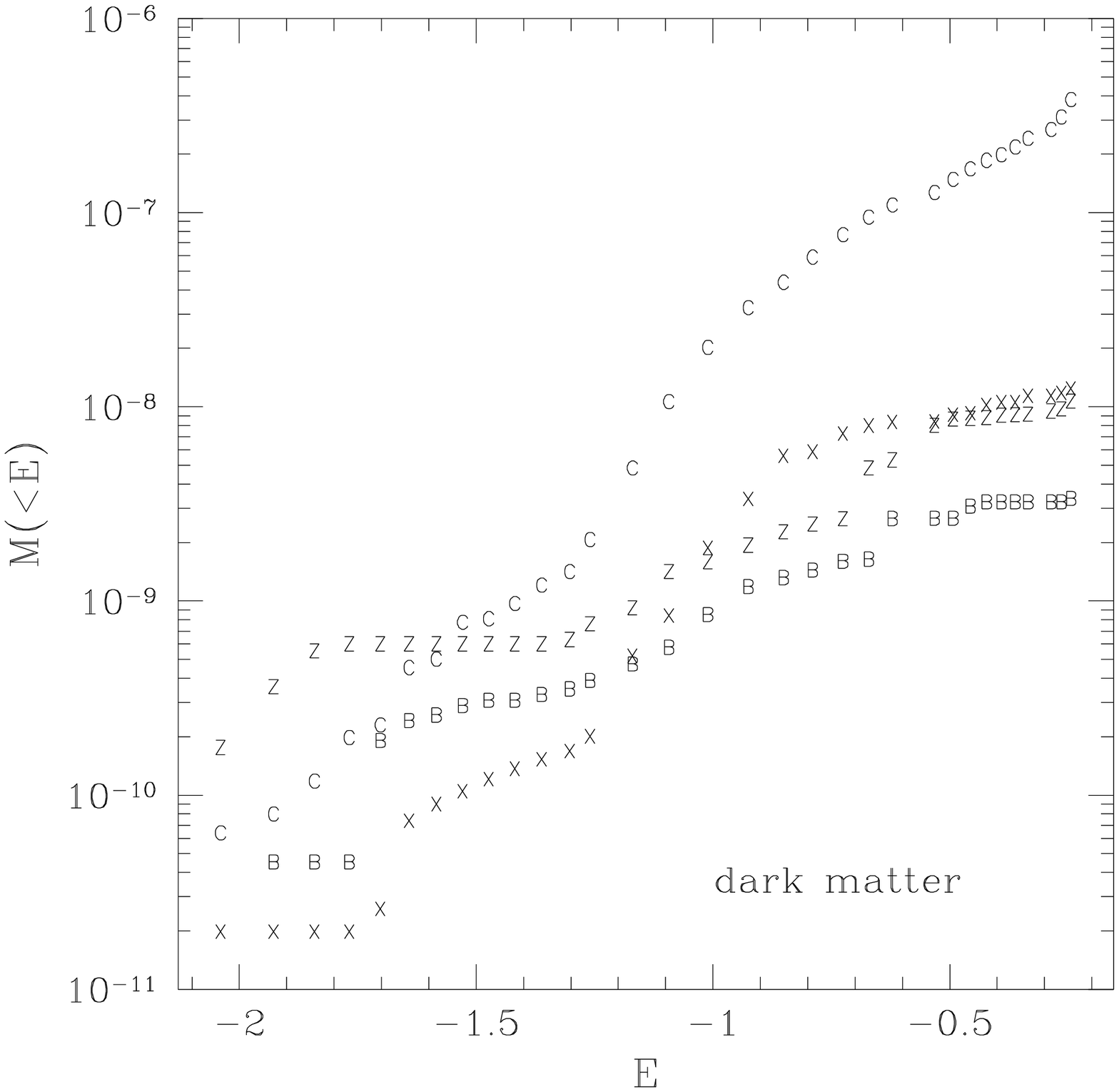} \\
              \end{array}$

  \caption{Cumulative, by mass, energy distributions of the various orbital families
in the self-consistent solution of  MOD1 (left column) and MOD1bis (right column).
The symbols "B," "X," "Z," and "C" denote the mass contributed by box, x-tube, z-tube,
and chaotic orbits, respectively.
  }\label{ME}
\end{figure}

\begin{figure}
$\begin{array}{cc}
                \includegraphics[width=3in,height=3in]{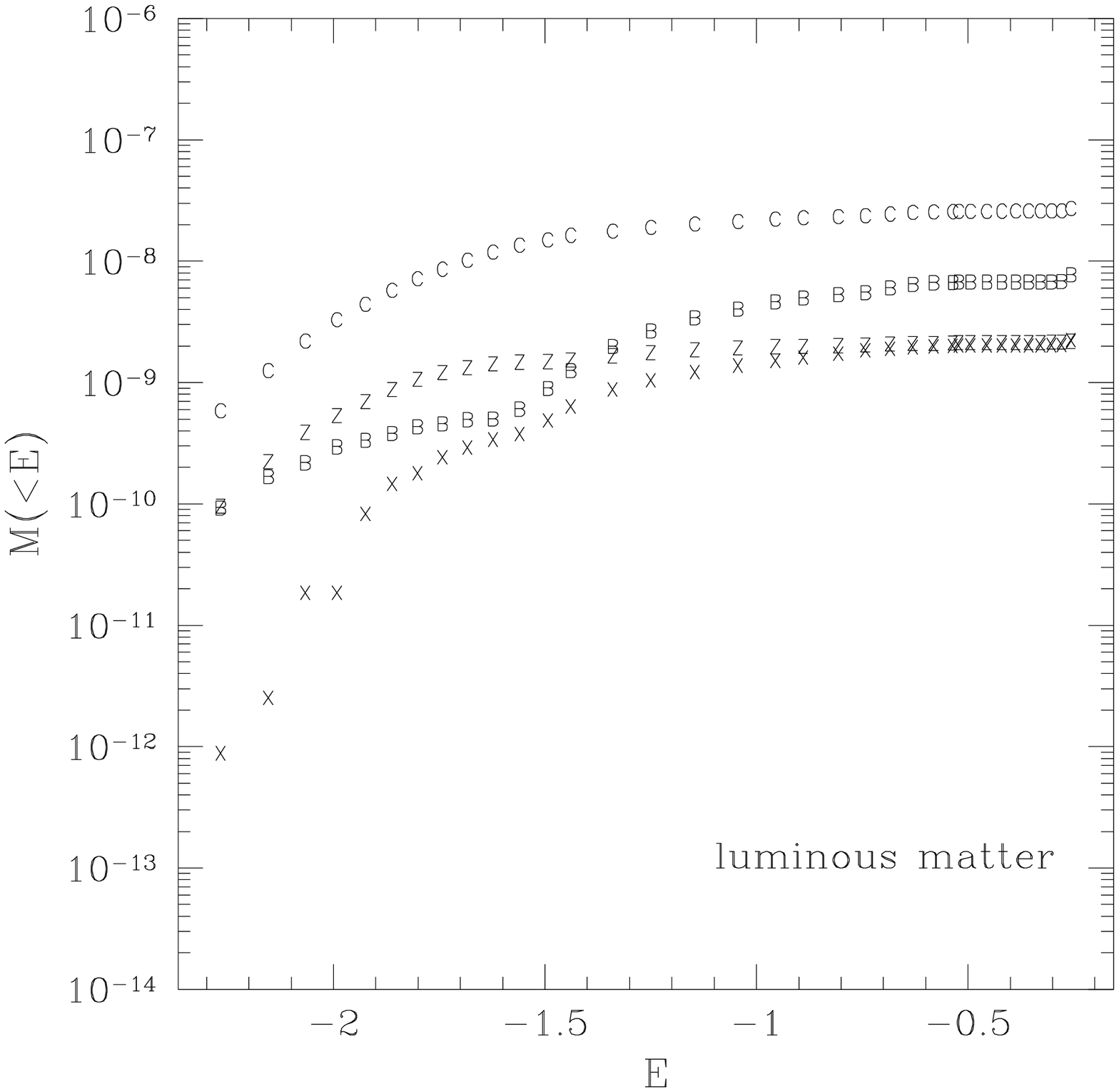}  &
 \includegraphics[width=3in,height=3in]{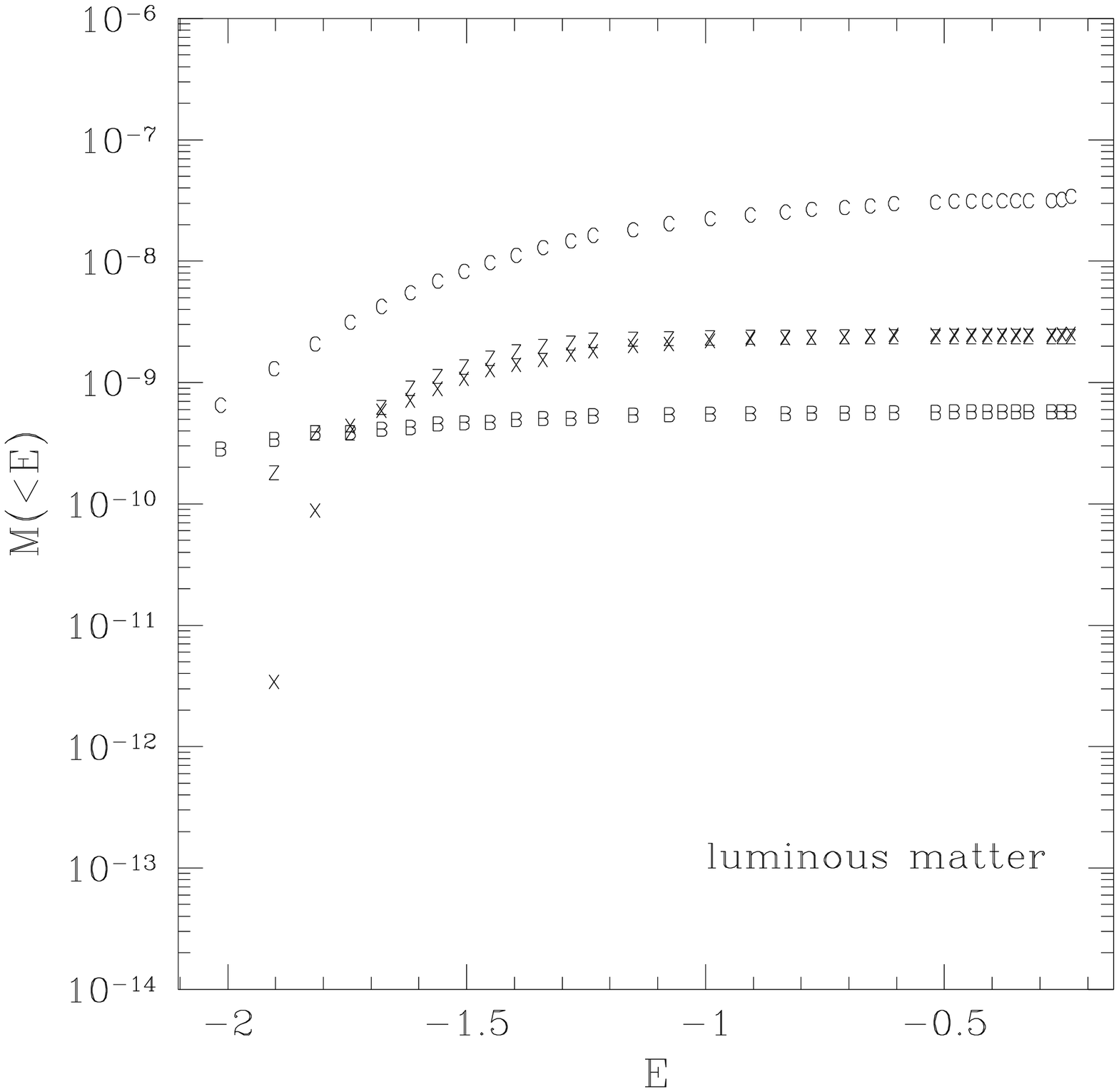} \\
                 \includegraphics[width=3in,height=3in]{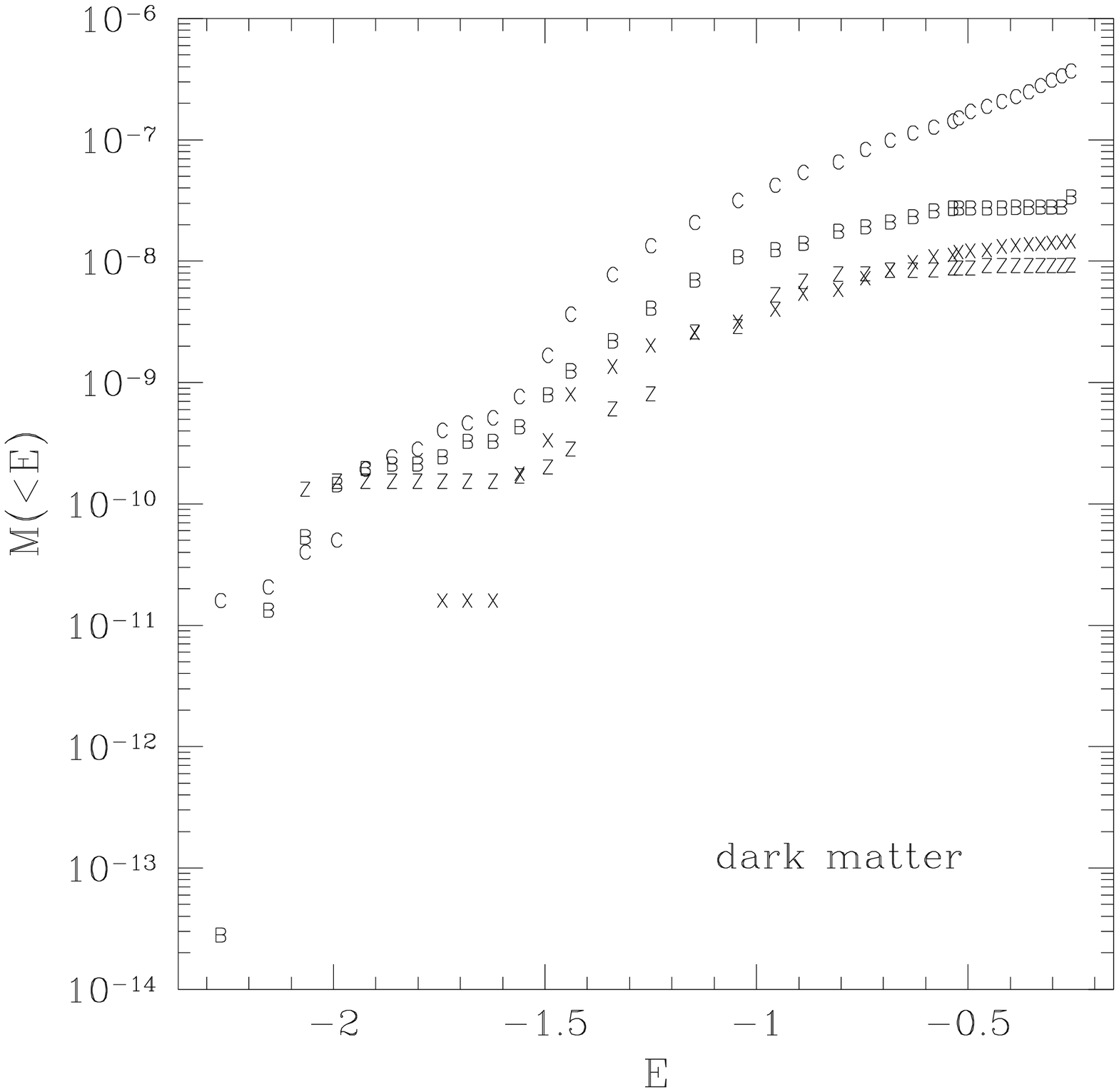} &
\includegraphics[width=3in,height=3in]{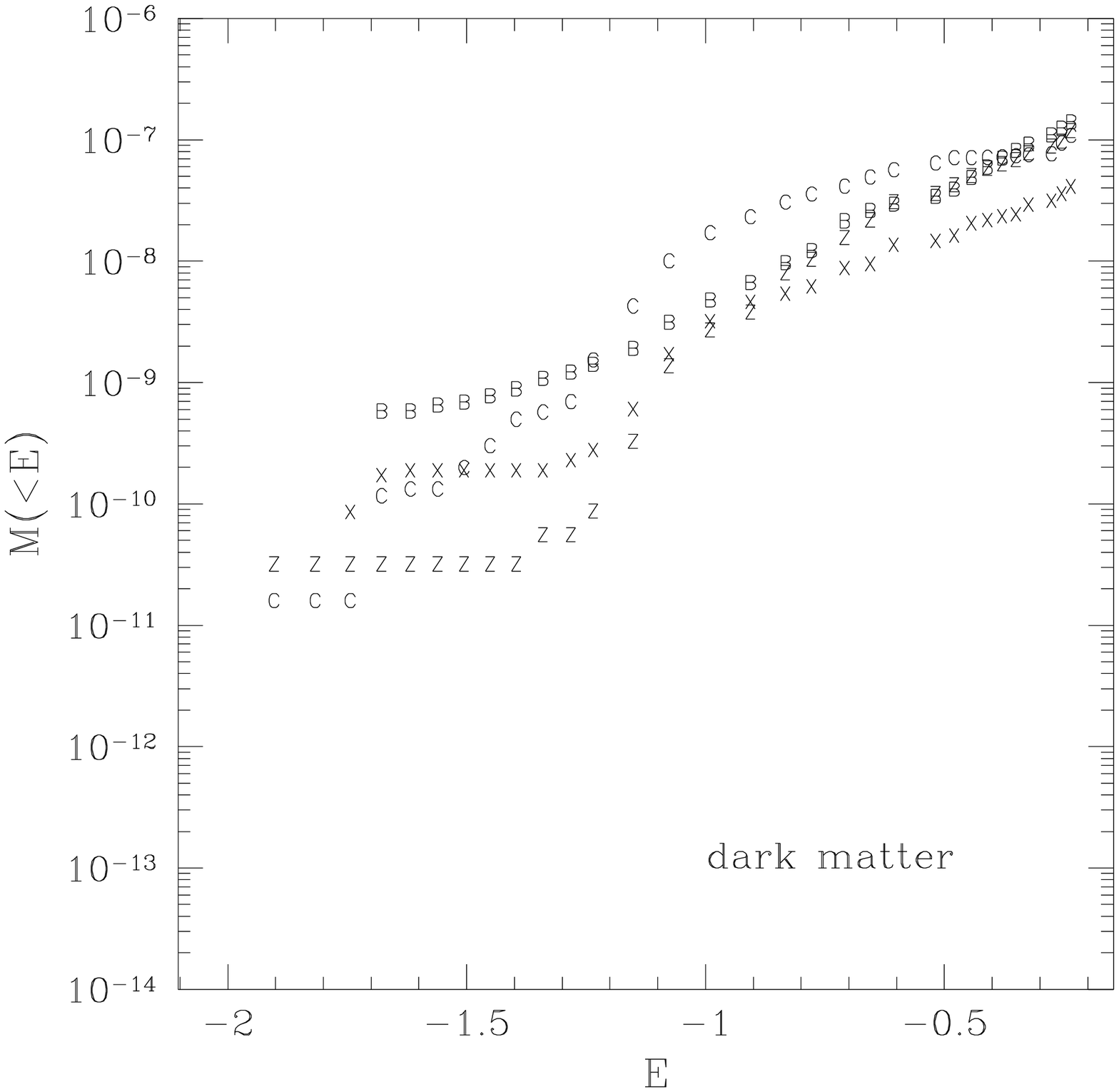} \\
              \end{array}$

  \caption{Cumulative, by mass, energy distributions of the various orbital families
in the self-consistent solution of  MOD2 (left column) and MOD3 (right column).
The symbols "B," "X," "Z," and "C" denote the mass contributed by box, x-tube, z-tube,
and chaotic orbits, respectively.
  }\label{ME2}
\end{figure}

\begin{figure}
  \includegraphics[width=6in,height=6in]{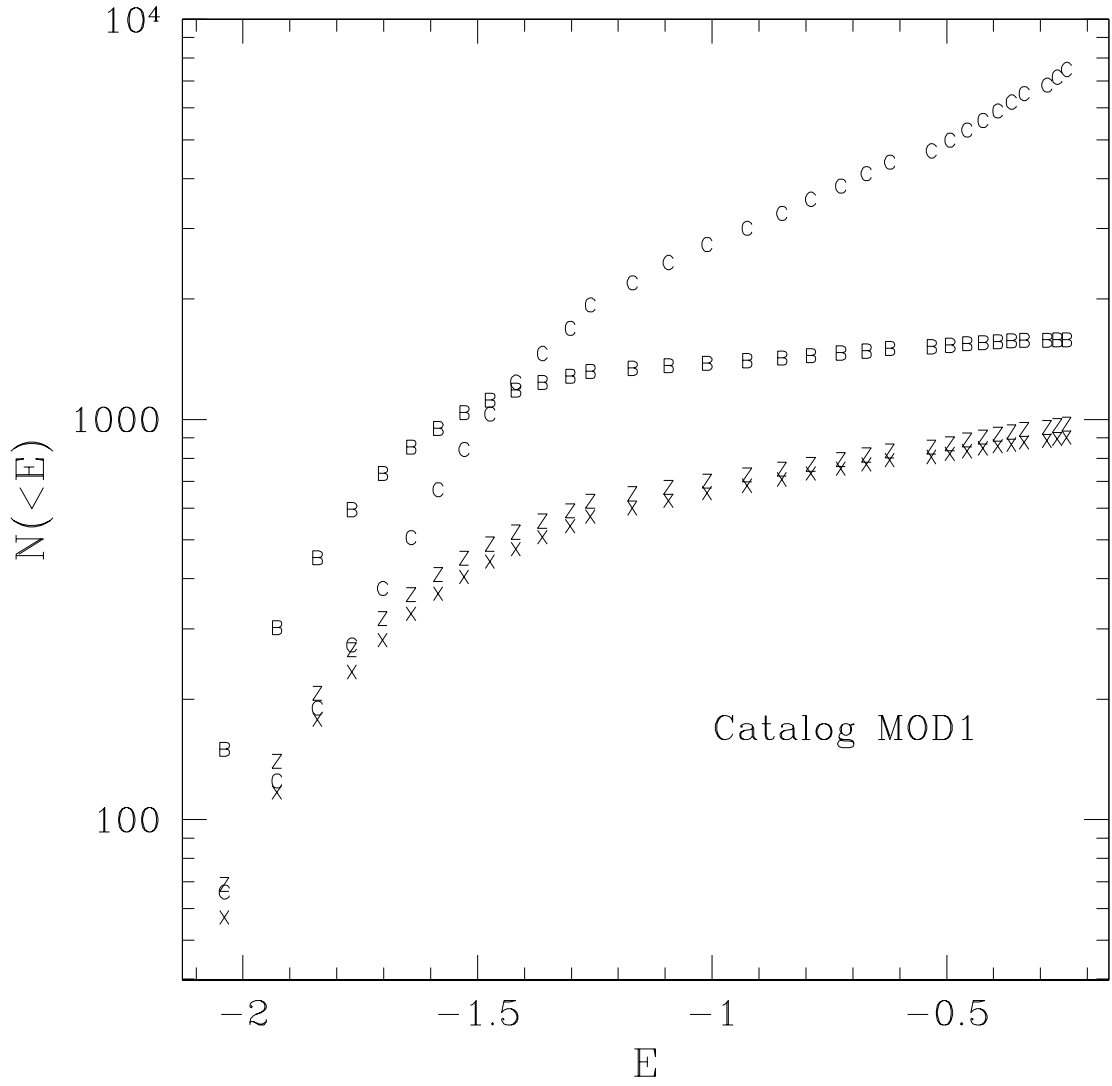}\\
  \caption{Cumulative, by number, energy distribution of the
  whole MOD1 orbital catalog. Symbols as in Fig. \ref {ME}}\label{NE}
\end{figure}

To give a finer detail on the self-consistent model orbital structure,
Fig. \ref{ME} and \ref{ME2} show the cumulative contribution, by mass, of the various orbital families
in our self-consistent solutions. Chaotic orbits are of increasing
importance at higher energies, especially for the dark matter component.
This is partly explained by the initial abundance of high energy chaotic orbits
in the full MOD1 catalog (see Fig. \ref{NE}).
Fig. \ref{NE} indicates, also, how the low-energy back-bone of the self consistent model
is mostly constituted by short-axis tube orbits; in the "more" evolved MOD1bis model
(described in more detail below), chaotic orbits of the luminous matter component are
important on the whole energy range.
All this suggests to investigate more deeply the role of orbital stochasticity,
as we do in Sect. 6.

\subsection{Velocity dispersion tensors}
To gain a more complete picture of the model kinematics,
we also evaluated the first and second velocity moments
in each of the self-consistent solutions.
Using the properties of symmetry with respect to the coordinate
planes in the cartesian frame, it was sufficient to store data for
any given orbit in only one of the eight octants.


We stored the velocities along an orbit symmetrizing them respect to
the principal axes in order
to insert the point in which one velocity is evaluated in a precise cell of 
the positive octant.
In this way an orbit, starting with a fixed direction, correspond to a trajectory 
covered equally (or similar) in two opposite directions. It results a mean motions in
each cell near to the zero value.
Then we weighted these data with the occupation number (mass) of the luminous orbits in each cell
obtaining:
$$
<V_{i}>_{lum}=\frac{\sum _{k=1}^{n_{orbit}}C_{k;lum}B_{k,j;lum}<v_{i}>_{lum}}{\sum
_{k=1}^{n_{orbit}}C_{k;lum}B_{k,j;lum}}
$$
and
$$
<V_{i}V_{h}>_{lum}=\frac{\sum
_{k=1}^{n_{orbit}}C_{k;lum}B_{k,j;lum}<v_iv_h>_{lum}}{\sum
_{k=1}^{n_{orbit}}C_{k;lum}B_{k,j;lum}}.
$$
Finally, we obtain the velocity dispersion tensor of the stars, $\sigma_{lum,ih}^2$ in
each cell $l$, as:
$$
\sigma_{lum,l;ih}^2 =<V_i V_h>_{lum,l}- <V_i>_{lum,l}<V_h>_{lum,l}.
$$

\begin{figure}
\centering
\includegraphics[width=6in,height=8.5in,angle=0]{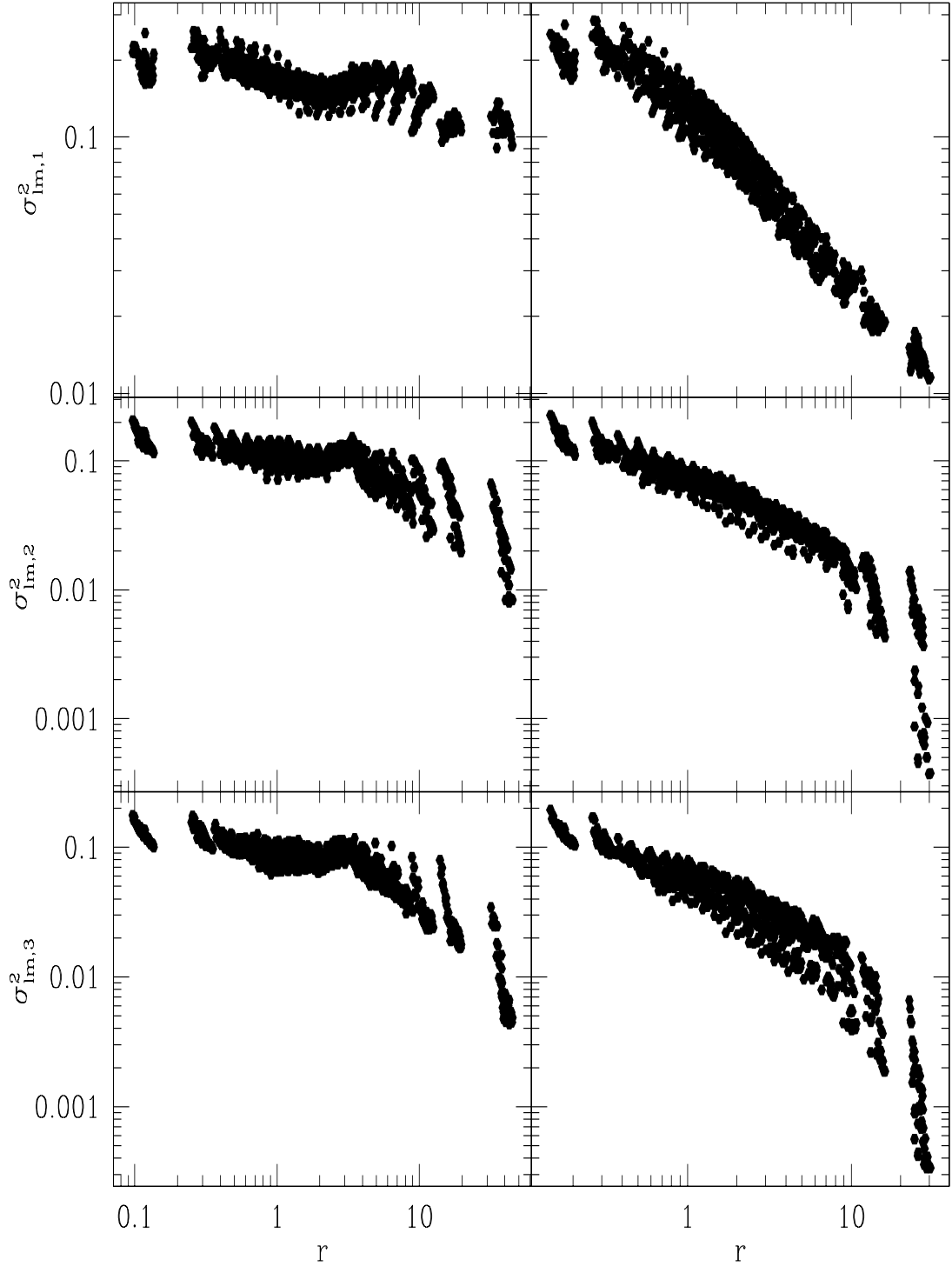}\\
\caption{The stellar principal velocity dispersions
${\sigma^2_{lum,l;i}}$, $i=1,2,3$ as functions of the distance
from the centre in model MOD1 (left column) and in a model with
the same luminous mass but without dark matter (right column).}
\label{sigmalum}
\end{figure}

The diagonalization of the velocity dispersion tensor, leads to the
three principal velocity dispersions, $\sigma_{lum,l;i}^{2}$
($i=1,2,3$) having the profiles shown in the left column of Fig.
\ref{sigmalum} (we order the dispersions
to have $\sigma_{lum,l;1}^2>\sigma_{lum,l;2}^2>\sigma_{lum,l;3}^2$).
In the right column of the figure we show, for comparison, the
corresponding principal dispersions evaluated in a model with only
the luminous component. As expected, the velocity dispersion in the
model with both luminous and dark matter have a flatter profile. The
difference in the external region is mainly due to insufficient
orbital sampling and the (necessary) cut-off in the dark matter
distribution.

\section{Importance of orbital stochasticity}
\label{discussione}
\begin{figure}
\centering
\includegraphics[width=6in,height=6in,angle=0]{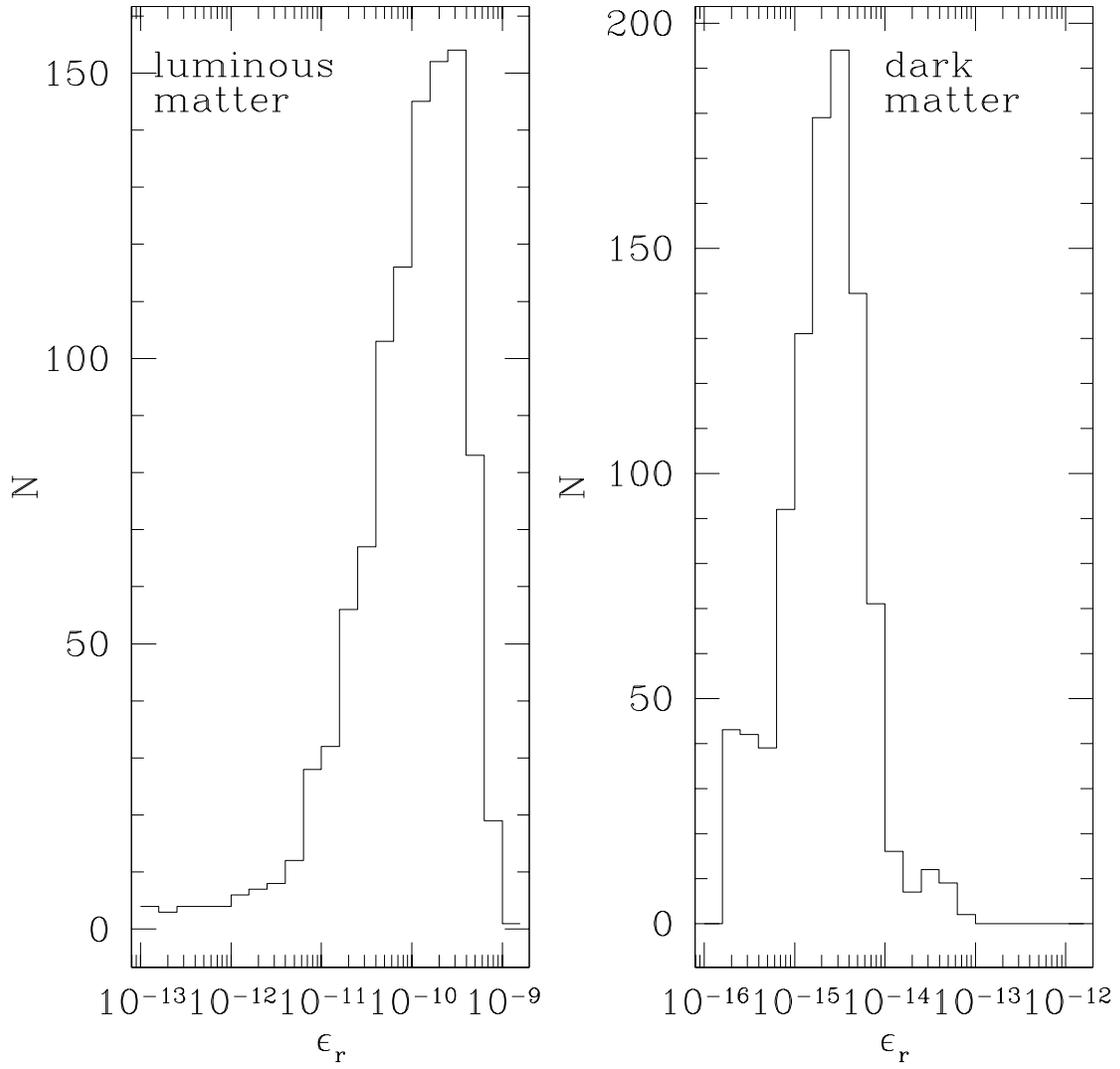}\\
\caption{Histograms of the relative errors $\epsilon_{r}$ for MOD1bis,
referring to the luminous and dark matter component, respectively. }\label{errori23}
\end{figure}

The Schwarzschild method for building self-consistent models
uses orbits as ``templates", with supposedly time-invariant
properties, so a requirement for a stationary solution is that
a large fraction of the orbits (whether regular or chaotic)
have nearly time-independent occupation numbers.
Using the method of Pfenniger discussed above, we found that,
at the time $2 T_H$, roughly $20 \%$ of the occupation
numbers of the orbits used to represent the luminous component
were not yet fixed in time, while for the dark matter this percentage
was as high as $\sim 50 \%$.
For MOD1, we attempted to obtain a more ``stationary" solution
by integrating the orbits up to $5 T_H$.
In this way, using a total of $2851$ orbits
for the luminous component and $2796$ for the dark matter,
we achieved a solution (MOD1bis) with a relative error in the mass
of each cell $< 10^{-9}$ (see Fig. \ref{errori23}).
In this new model, the number of orbits with non-steady $B_{k,j}$
values was reduced to $\sim 15 \%$, which would seem to indicate the
presence in the initial solution of
quite a few ``sticky'' chaotic orbits.

 \begin{table} \small
 \begin{center}
\begin{tabular}{lcccc}
\multicolumn{5}{c}{\textbf{TABLE 2}} \\
  \hline
  \hline
 \multicolumn{1}{l}{Set} & \multicolumn{2}{c}{Luminous matter}& \multicolumn{2}{c}{Dark matter}
  \\
  \multicolumn{1}{l}{} & \multicolumn{1}{c}{$\langle w\rangle$}& \multicolumn{1}{c}{$\sigma_{\langle w\rangle}$} &\multicolumn{1}{c}{$\langle w\rangle$}& \multicolumn{1}{c}{$\sigma_{\langle w\rangle}$}\\
  \hline
 \multicolumn{1}{l}{X} & \multicolumn{1}{c}{$0.000872$}& \multicolumn{1}{c}{$0.00138$} &\multicolumn{1}{c}{$0.00388$}& \multicolumn{1}{c}{$0.00137$}\\
  \hline
   \multicolumn{1}{l}{Y} & \multicolumn{1}{c}{$-0.00845$}& \multicolumn{1}{c}{$0.00150$} &\multicolumn{1}{c}{$-0.01$}& \multicolumn{1}{c}{$0.00165$}\\
  \hline
   \multicolumn{1}{l}{Z} & \multicolumn{1}{c}{$0.00929$}& \multicolumn{1}{c}{$0.00177$} &\multicolumn{1}{c}{$0.00771$}& \multicolumn{1}{c}{$0.00157$}\\
  \hline

\end{tabular}
 \caption{The mean values of the "evolution" parameter $w$ and their standard deviations.}\label{tab2}
\end{center}
\end{table}

Evaluating stochasticity based on orbital occupation numbers
is prone to error, since even regular orbits can take
long times to fill their invariant tori.
To more accurately identify stochastic orbits,
we used the Smaller Alignment Index (SALI) introduced by
Skokos (2001), and
later applied successfully to different dynamical systems (Skokos et
al. 2003; Skokos et al. 2004; Manos \& Athanassoula 2006). It is
defined as (Skokos 2001)
$$
SALI(t)\!=\!min\! \Bigg\{\! \Bigg\|\! \frac{\textbf{w}_1(t)}{\| \textbf{w}_1(t) \|}\! + \!\frac{\textbf{w}_2(t)}{\| \textbf{w}_2(t) \|}\! \Bigg\|\!,\!\Bigg\| \!\frac{\textbf{w}_1(t)}{\| \textbf{w}_1(t)\|}\! -\! \frac{\textbf{w}_2(t)}{\| \textbf{w}_2(t)\|}\! \Bigg\| \! \Bigg\}
$$
where $\textbf{w}_1(t)$ and $\textbf{w}_2(t)$ are two deviation
vectors in phase space, centered in the initial condition of one
orbit and pointing in two different arbitrary directions.
In the case of
Hamiltonian flows, SALI fluctuates around a non-zero value for
regular orbits, while it tends (in time) to zero for chaotic orbits.



\begin{figure}
\begin{center}
\includegraphics[width=6in,height=6in]{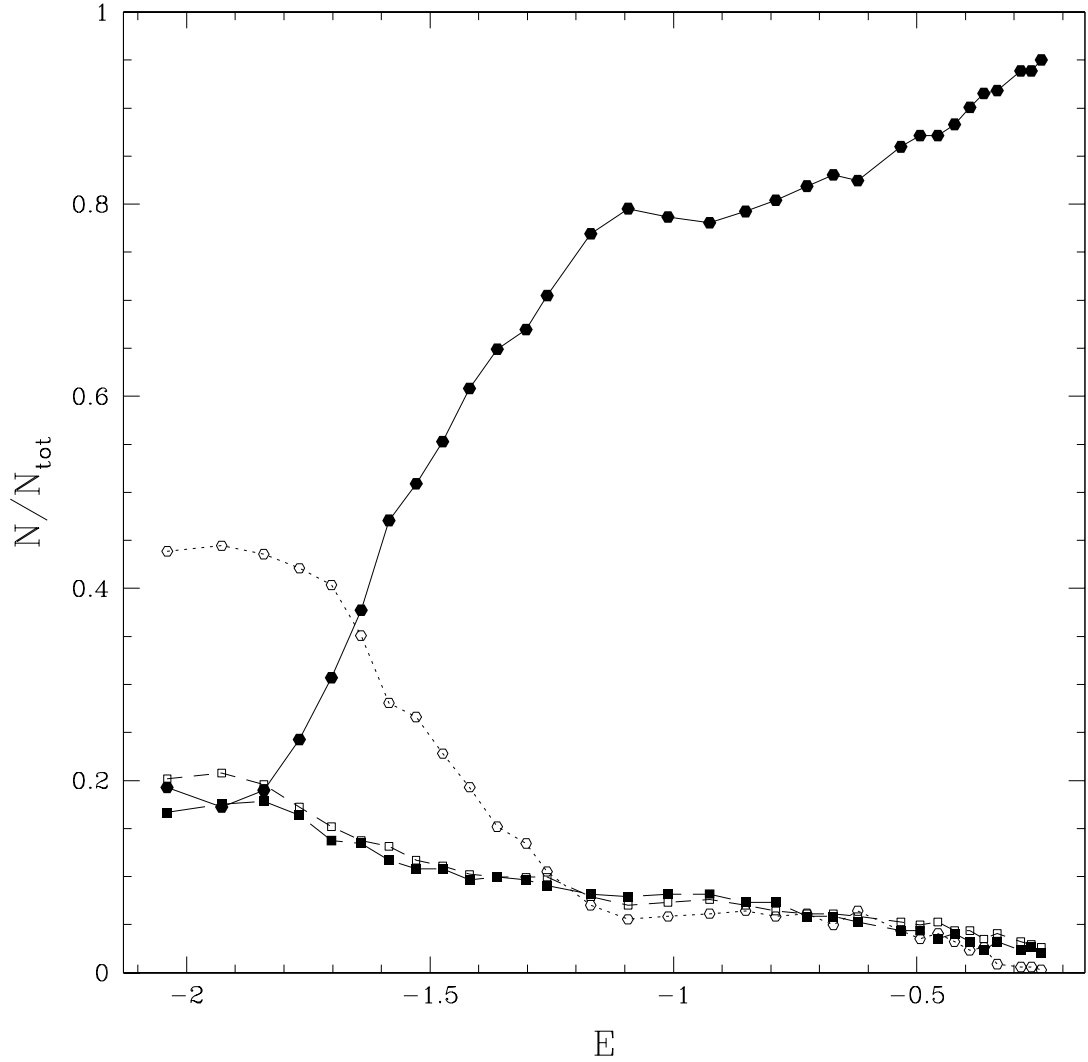}\\
\caption{Fraction of different types of orbits in the full MOD1
orbital catalog vs energy. Full circles represent stochastic orbits;
open circles are box-like orbits; open and full
squares correspond to short and outer tube orbits respectively.}
\label{entipo}
\end{center}
\end{figure}

By studying the evolution of this quantity for all the orbits of the catalog used
to seek for the self-consistent solution of MOD1
we found that $68 \%$ of them are chaotic, especially at
high energies (see Fig. \ref{entipo}). Regarding the orbits of the
subsample giving the self-consistent solution MOD1bis, we found
that more than $90\%$ of non ergodic orbits are chaotic.

\begin{figure}
\begin{center}
\includegraphics[width=6in,height=6in]{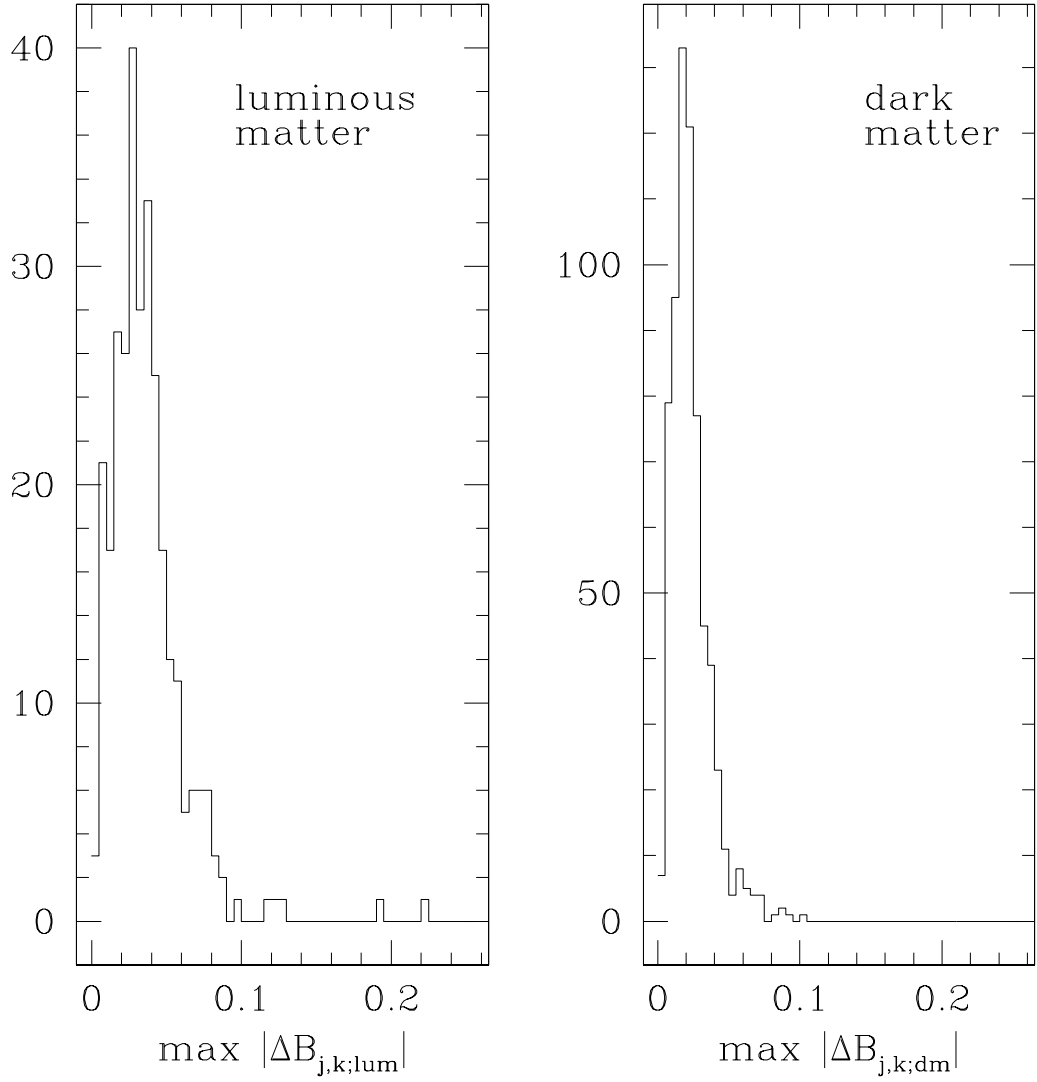}\\
\caption{Histograms of the maximum $|\Delta B_{j,k}|$  for the stochastic orbits used in MOD1bis
integrated once up to $5 T_H$ and once up to $6 T_H$. }
\label{diffbij}
\end{center}
\end{figure}

To evaluate the degree to which the stochastic orbits in MOD1bis
would cause this model to evolve over long periods,
we recorded the position and velocity of every chaotic
orbit at the end ($t=5 T_H$) of the integration and used these
values as initial conditions for a new integration over another $T_H$.
We then again recorded the occupation numbers and compared
them with the old ones. As shown in Fig. \ref{diffbij}, more than
$90 \%$ of these ``reintegrated" orbits have fluctuations in
$B_{i,j}$ greater than the canonical value of $0.005$. Operating in a similar
way for dark matter, we reached the conclusion that our self-consistent
model contains $\sim 10 \%$ and $\sim 23 \%$ of still-evolving
stellar and dark matter orbits, respectively.

\begin{figure}
\centering
\includegraphics[width=6in,height=8.5in]{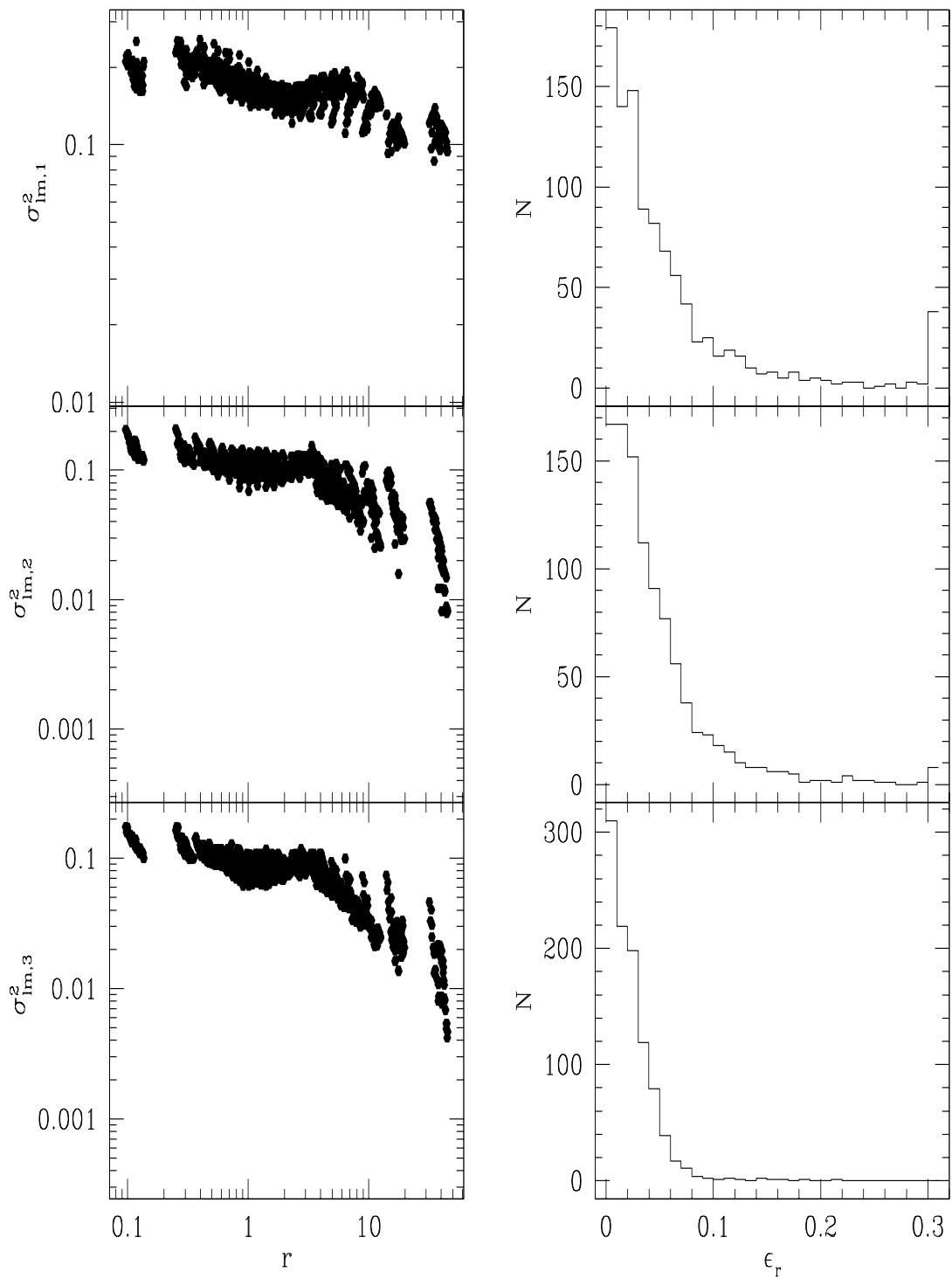}
\caption{Left column: principal velocity dispersions for the
luminous component, obtained with the self-consistent solution realized
after having integrated orbits over $5 T_H$ (MOD1bis).
Right column:
relative change in this quantity with respect to that
obtained with the self-consistent solution after having
integrated the orbits over $2 T_H$.}
\label{diffdisp}
\end{figure}

In order to study the cell density change as a result of orbital evolution,
and so the role of evolution on the overall shape of the model galaxy,
we studied the variation of the occupation numbers of the chaotic-non ergodic orbits
on a $T_H-5T_H$ time basis, of three set of cells (named $X,Y,Z$), sampling the volumes
around the three coordinate axes. At this scope we defined, in each cell ($i$),
an "evolution" parameter
\begin{equation}\label{ordinata}
\displaystyle
    w_i =\frac{\sum_{k\in K1}B_{i,k}|_{5th} - \sum_{k\in K1}B_{i,k}|_{1th}
    }{\sum_{k\in K2}B_{i,k}|_{5th}}
\end{equation}
where in the numerator summation is extended over all the
chaotic-non ergodic orbits in the self-consistent solutions and, to normalize,
we divide for the sum over the whole set of orbits in the self-consistent solution
at $t=5 T_H$. The average value of $w$ on the $X,Y,Z$ volumes gives an indication of the
shape evolution: a globally unchanged shape would correspond to
$\langle w\rangle_{X}=\langle w\rangle_{Y}=\langle w\rangle_Z=0.$
We found that for luminous matter $<w>_X$ is compatible with being $=0$, while
$\langle w\rangle_Y<0$ and, $\langle w\rangle_Z>0$. This corresponds to a slight 
evolution toward a more prolate shape. The dark matter component shows a, statistically 
significant, increase in time of the chaotic orbit population of both the $X$ and $Z$ 
volumes, and a (smaller) depopulation
of the $Y$ volume. This should corresponds, too, toward a more prolate configuration.
In view of future astrophysical applications (e.g. to the study of orbital the
evolution of massive black holes or globular clusters subjected to dynamical
friction), it is relevant to check how much the kinematical structure of
the galactic models depend on the presence of a fraction of evolving orbits.
To do this, we compared the principal velocity dispersions in MOD1bis with
those of MOD1.
As we can see by comparing Fig. \ref{diffdisp} with Fig. \ref{sigmalum},
the behaviours are similar, with a relative difference that is almost
always below $10\%$.
We checked that the largest relative differences refer to the outermost shell.



\section*{Acknowledgments}
D.M. acknowledges support from grants AST-0206031, AST-0420920 and
AST-0437519 from the NSF, grant NNG04GJ48G from NASA, and grant
HST-AR-09519.01-A from STScI.  This work was supported in part by
the Center for Advancing the Study of Cyberinfrastructure at the
Rochester Institute of Technology.

\appendix
\section{Potential and forces}
\label{potcons}

Using the substitution $s=(1+\tau)^{-1/2}$ in Eq. \ref{potl} and
\ref{potdm}, they transforms into:
\begin{equation}\label{potls}
\Phi_{l}(\textbf{x})=\displaystyle -GM\int_{0}^{1}\frac{ds}{\Delta'}
\left[ \frac{1}{(1+m)^2}\right]
\end{equation}

\begin{eqnarray}
\label{potdms}\displaystyle \lefteqn{\Phi_{dm}(\textbf{x})=-2\pi
Ga_{dm}b_{dm}c_{dm}\rho_{dm,0}{}} \nonumber\\ && \! \! \!
\!\!\!\times\!\! \int_{0}^{1} \!\!\frac{ds}{\Delta'} \!\left(\!
\frac{\pi}{2}\!-\! \textrm{atan} m'\!+\!\log\!
\left(\!1\!+\!m'\!\right)\!-\!\frac{1}{2}\!\log \!\left(\!1\!+\!
m'^{2}\! \right)\!\right),
\end{eqnarray}
where
\begin{equation}
\label{deltas}
\begin{array}{ll}\displaystyle\!\!\!\!\!
m^{2}(s)\!\!=\!\!s^{2}\!\left[\!\frac{x^{2}}{1+s^{2}({a_l}^{2}-1)}\!+\!\frac{y^{2}}{1+s^{2}({b_l}^{2}-1)}\!+\!\frac{z^{2}}{1+s^{2}({c_l}^{2}-1)}\!\right] \\ \\
\!\!\!\!\!\tilde{\Delta}\!
=\!\sqrt{[1+s^{2}({a_l}^{2}-1)][1+s^{2}({b_l}^{2}-1)][1+s^{2}({c_l}^{2}-1)]}
\end{array}
\end{equation}
and
\begin{equation}
\label{delta1s}
\begin{array}{ll}\displaystyle\!\!\!\!\!
m'^{2}(s)\!\!=\!\!s^{2}\!\left[\!\frac{x^{2}}{1\!+\!s^{2}\!(\!{a_{dm}}^{2}\!-\!1\!)}\!+\!\frac{y^{2}}{1\!+\!s^{2}\!(\!{b_{dm}}^{2}\!-1)\!}\!+\!\frac{z^{2}}{\!1\!+\!s^{2}\!(\!{c_{dm}}^{2}\!-1\!)}\!\right]\! \\ \\
\!\!\!\!\!\tilde{\Delta'}\!\!
=\!\!\sqrt{\![1+s^{2}\!({a_{dm}}^{2}-1)\!]\![1+s^{2}\!({b_{dm}}^{2}-1)\!]\![1+s^{2}\!({c_{dm}}^{2}-1)\!]}\!.
\end{array}
\end{equation}

With the substitutions $s=a_{l;i}(a_{l;i}^{2}+\tau)^{-1/2}$ for the
luminous forces and $s=a_{dm;i}(a_{dm;i}^{2}+\tau)^{-1/2}$ for that
of dark matter we transform Eq. \ref{forcel} and \ref{forcedm} in:

\begin{equation}
\label{forcels}\displaystyle F_{l_{i}}=-GM\frac{\partial
\Phi_{l}}{\partial
x_{i}}=-2x_{i}\int_{0}^{1}\frac{s^{2}ds}{\Delta_{i}m_{i}(1+m_{i})^3}
\end{equation}

\begin{eqnarray}
\label{forcedms}\displaystyle \lefteqn{F_{dm_{i}}=-\frac{\partial
\Phi}{\partial x_{i}}=-4\pi Ga_{dm}b_{dm}c_{dm}\rho_{dm,0}x_{i}{}
}\nonumber\\ & & \quad \quad \times \!\!
\int_{0}^{1}\frac{s^{2}ds}{\Delta'_{i} (1+m'_{i})(1+{m'_{i}}^{2})}
\end{eqnarray}

where $x_1=x$, $x_2=y$, $x_3=z$ and:
\begin{equation}
\label{deltasforce}\displaystyle
\begin{array}{ll}
{m_{i}}^{2}(s)=s^{2}\left[\frac{x^{2}}{{a_{l;i}}^{2}+C_{l;1}s^{2}}+\frac{y^{2}}{{a_{l;i}}^{2}+C_{l;2}s^{2}}+\frac{z^{2}}{{a_{l;i }}^{2}+C_{l;3}s^{2}}\right] \\ \\
{m'_{i}}^{2}(s)\!=\!s^{2}\!\left[\!\frac{x^{2}}{{a_{dm;i}}^{2}\!+\!C_{dm;1}\!s^{2}}\!+\!\frac{y^{2}}{{a_{dm;i}}^{2}\!+\!C_{dm;2}\!s^{2}}\!+\!\frac{z^{2}}{ {a_{dm;i }}^{2}\!+\!C_{dm;3}\!s^{2}}\!\right]\! \\ \\
\Delta_{i}
=\sqrt{({a_{l;i}}^{2}+A_{l;1}s^{2})({a_{l;i}}^{2}+A_{l;2}s^{2})({a_{l;i}}^{2}+A_{l;3}s^{2})} \\ \\
\Delta'_{i}
\!=\!\sqrt{\!({a_{dm;i}}^{2}\!+\!A_{dm;1}\!s^{2}\!)\!(\!{a_{dm;i}}^{2}\!+\!A_{dm;2}\!s^{2}\!)\!(\!{a_{dm;i}}^{2}\!+\!A_{dm;3}\!s^{2}\!)}.
\end{array} \end{equation}

The costants in Eq. \ref{deltasforce} are
\begin{equation}
\label{const}\displaystyle
\begin{array}{ll}
i=1:\\
\quad A_{j;1}={b_j}^{2}-1 \qquad A_{j;2}={c_j}^{2}-1\qquad A_{j;3}={a_j}^{2}\\
\quad C_{j;1}=0\qquad C_{j;2}={b_j}^{2}-1\qquad C_{j;3}={c_j}^{2}-1\\
i=2:\\
\quad A_{j;1}={c_j}^{2}-{b_j}^{2} \qquad A_{j;2}=1-{b_j}^{2} \qquad A_{j;3}={b_j}^{2}\\
\quad C_{j;1}=1-{b_j}^{2} \qquad C_{j;2}=0 \qquad C_{j;3}={c_j}^{2}-{b_j}^{2}\\
i=3:\\
\quad A_{j;1}=1-{c_j}^{2} \qquad A_{j;2}={b_j}^{2}-{c_j}^{2} \qquad A_{j;3}={c_j}^{2}\\
\quad C_{j;1}=1-{c_j}^{2} \qquad C_{j;2}={b_j}^{2}-{c_j}^{2} \qquad C_{j;3}=0.\\
\end{array}
\end{equation}
with $j=l$ or $j=dm$ if we deal with luminous or dark matter, respectively.


\label{lastpage}

\end{document}